\documentclass[twocolumn,eqsecnum,pre,showpacs,floatfix]{revtex4-1}%
\usepackage{graphicx}
\usepackage{amsmath}
\usepackage{bm}
\usepackage{epsfig}
\usepackage{amsfonts}
\usepackage{amssymb}
\usepackage[usenames]{color}
\begin{document}

\newcommand{\Ochange}[1]{{\color{red}{#1}}}
\newcommand{\Ocomment}[1]{{\color{PineGreen}{#1}}}
\newcommand{\Hcomment}[1]{{\color{ProcessBlue}{#1}}}
\newcommand{\Hchange}[1]{{\color{BurntOrange}{#1}}}

\title{The collapse transition of randomly branched polymers --
 renormalized field theory}
\author{Hans-Karl Janssen}
\affiliation{Institut f\"ur Theoretische Physik III, Heinrich-Heine-Universit\"at, 40225
D\"usseldorf, Germany}
\author{Olaf Stenull}
\affiliation{Department of Physics and Astronomy, University of Pennsylvania, Philadelphia
PA 19104, USA}
\date{\today}

\begin{abstract}
\noindent We present a minimal dynamical model for randomly branched isotropic polymers,
and we study this model in the framework of renormalized field theory. For the
swollen phase, we show that our model provides a route to understand the well
established dimensional-reduction results from a different angle. For the
collapse $\theta$-transition, we uncover a hidden Becchi-Rouet-Stora super-symmetry, signaling the sole relevance of tree-configurations.
We correct the long-standing 1-loop results for the critical exponents, and we
push these results on to 2-loop order. For the collapse $\theta^{\prime}$-transition, we find a runaway of the renormalization group flow, which lends
credence to the possibility that this transition is a fluctuation-induced
first-order transition. Our dynamical model allows us to calculate for the first time the fractal dimension of the shortest path on randomly branched polymers in the swollen phase as well as at the collapse transition and related fractal dimensions.

\end{abstract}
\maketitle

\section{Introduction}
\label{sec:introduction}

Randomly branched polymers (RBPs) are a classical topic in statistical
physics. Seminal advancements in the theoretical understanding of these
polymers~\cite{LuIs78,IsLu80,HaLu81,PaSo81,Sha83} have been made not long after the
advent of renormalization group theory starting with the seminal work of Lubensky and Issacson (LI). With the surge of biophysics, there recently has been renewed interest in RBPs because RNA in its molten phase belongs to the same universality class as swollen RBPs~\cite{deGennes1968,bundschuhTwa1999}.
However, the current understanding of RBPs is still not quite satisfactory. For example, the topology of their phase diagram is not entirely clear. In particular the part of the phase diagram that contains the so-called $\theta^{\prime}$-transition gives reason for debate. The existing theories~\cite{LuIs78,IsLu80} for the collapse $\theta$-transition are not entirely correct. As far as we know, there exist no theories for the transport properties and the related fractal dimensions of RBPs such as the dimensions of the backbone, the shortest path and so on.

In this paper we are not interested in chemical or mechanical
properties of randomly branched polymers. Rather, we are interested in their structure. More precisely, we are interested in their universal structural properties in the limit where the number of constituent monomers is large. In this limit, an RBP can be regarded as a large cluster, and its structural properties are universal, i.e., common to large RBPs as a class irrespective of their physical or chemical details. Phenomenologically, only
their large size and their branching on all length scales are relevant. In the
language of critical phenomena -- phenomena with large correlation lengths,
here the diameters of clusters -- all such systems of fractal clusters with different
microscopic aspects but with these common relevant properties belong to one
universality class, which we denote in the following with the \emph{pars pro toto}
randomly branched polymers. In computer simulations such clusters are
usually constructed as so-called lattice animals, i.e.,  clusters of connected sites (monomers) on a $d$-dimensional regular lattice. The recent publication of Hsu
and Grassberger on the collapse transition of animals \cite{HsGr05} and the unresolved issues mentioned above have triggered us to reconsider this classical topic with field theoretic methods.

In the much-studied case of a single large linear polymer in a diluted solvent, the phase diagram is one-dimensional. When the solvent quality is lowered (typically by lowering its temperature) below the so-called $\theta$-point, the polymer undergoes a collapse transition from a swollen coil-like conformation to a compact globule-like conformation. In simple lattice models, the monomer-solvent repulsion that drives the collapse transition is generically implemented via an effective attractive interaction between non-bonded monomers which is equivalent to the monomer-solvent repulsion at least as far as universal properties are concerned. Thus, the fugacity for non-bonded monomer-monomer contacts, let's call it $z_{\text{cont}}$, can be chosen as the control variable spanning the phase diagram of a linear polymer in a solvent. Evidently, $z_{\text{cont}}$ is closely related to temperature. 

\begin{figure}[ptb]
\includegraphics[width=6.5cm]{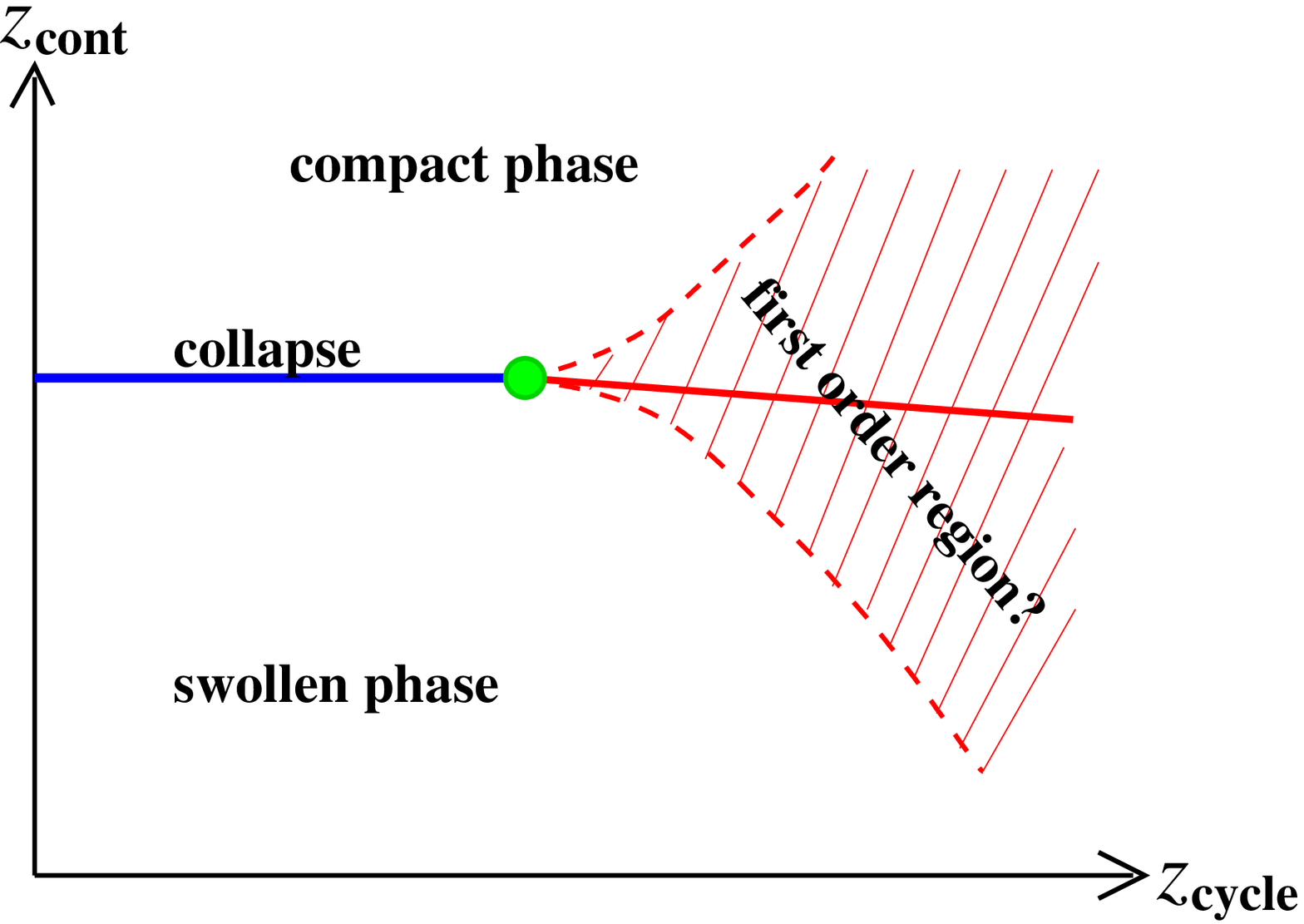}\caption{(Color online) Schematic phase diagram
for an RBP modeled by a lattice animal in the limit of a large number of constituents. $z_{\text{cont}}$ is the fugacity for contacts between non-bonded monomers, and $z_{\text{cycle}}$ is the fugacity for closed monomer cycles on the polymer.}
\label{fig:phaseDia}
\end{figure}

In the case of a single large RBP in a diluted solvent, the phase diagram is two-dimensional, see Fig.~\ref{fig:phaseDia}. The basic reason for the additional dimension is that one has to deal with an additional fugacity stemming from the fact that the number of bonds $b$ of an RBP is not uniquely determined by its number of sites $N$, $b-N+1 =: l \geq 0$, whereas it is uniquely determined for a linear polymer (as well as for a tree-like branched polymer) with $l=0$. The additional fugacity, let's call it $z_{\text{cycle}}$, regulates thencyclomatic index (the number of cycles $l$) of the polymer in the grand partition sum. For $z_{\text{cycle}} = 0$, the RBP has no cycles and the minimal number of bonds, i.e., it is tree-like. The phase diagram becomes one-dimensional (it reduces to the vertical axis with $z_{\text{cycle}}=0$ in Fig.~\ref{fig:phaseDia}). Physically, $z_{\text{cycle}}$ can be varied, e.g., by adding polyfunctional chemical units to the solution whose insertion into RBP results in additional bond cycles.

Over the last two decades or so, a number of numerical studies have been undertaken to map out this phase diagram~\cite{HsGr05,HsNaGr05,DeHe83,HeSe96,SeVa94,FGSW92/94,JvR97/99/00}. The picture that arises from these studies can be summarized as follows:
There is a swollen phase where the
polymer is in a tree-like or sponge-like conformation and a compact phase,
where the polymer is in a coil-like or vesicle-like conformation. There is
some debate, whether there exists a phase transition between the two compact
conformations or not. Between the swollen and the compact phases, there is a
line of collapse transitions. One part, called the
$\theta$-line (labelled collapse, blue), corresponds to continuous transitions with universal
critical exponents from the tree-like conformation to the coil-like
confirmation. The other part of the transition line, called the $\theta
^{\prime}$-line (red, in the dashed region), corresponds to the transition
between the foam- or sponge-like conformation to the to vesicle-like
conformation. Between the $\theta$- and $\theta
^{\prime}$-lines there is a tricritical point. There has been some controversy, if the $\theta
^{\prime}$ transition is continuous or not. With the assumption of it being continuous, computer
simulations in 2 dimensions yield nonuniversal critical
exponents~\cite{HsGr05}. As we will explain in detail below, our RG study
shows that the collapse transition to the right of  the tricritical point is characterized by a runaway of the RG
flow. This suggests that the $\theta^{\prime}$-transition is a fluctuation induced first order transition instead.
It could also mean that two of the lines observed in numerical studies of the phase diagram, viz.\ the lines interpreted as the line of transitions between two compact phases and the $\theta^{\prime}$-line, respectively, are merely shadows of the spinodals of the discontinuous transition.

The most fruitful theoretical approach to RBPs is based on the asymmetric
Potts model \cite{LuIs78,HaLu81,Con83} although Flory theory \cite{IsLu80} and
real space renormalization \cite{FaCo80} have also been applied successfully.
For the swollen phase, the field theoretic problem was settled by Parisi and
Sourlas (PS) via mapping the relevant part of the asymmetric Potts model to
the Yang-Lee edge problem using dimensional reduction \cite{PaSo81}. Subsequently, this mapping has been applied to further problems such as the exact calculation of universal scaling functions characterizing the behavior in the physical dimension 3 \cite{JaLy92,JaLy94,Ca01}.
Dimensional reduction was confirmed later with the discovery of an exact relationship
between swollen RBP-models and repulsive gases at negative activity in two
fewer dimensions by Brydges and Imbrie \cite{BrIm03,Ca2003}.

The asymmetric Potts model also provides a vantage point for studying the $\theta$-transition and
is the basis of the seminal field theoretic work of LI
\cite{LuIs78} and Harris and Lubensky \cite{HaLu81}. Their $1$-loop
calculation for the $\theta$-transition, however, contains a systematic error
in the RG procedure, and as a consequence their long-standing $1$-loop results
for the collapse transition are strictly speaking not correct although the
numerical deviation from the correct results is fortunately small.

Very recently, we developed a new dynamical field theory for RBPs,
see~\cite{janssenStenull2010} for a brief account. In the present paper, we extend our work, and we present it in more detail to make it easier accessible for non-specialist readers. Our theory is based on a stochastic epidemic process which models especially dynamical percolation
with a tricritical instability \cite{JaMuSt04}. As we will discuss in detail below, we focus entirely on the the non-percolating phase of the process. There, the very large clusters that finally result have the same statistics as RBPs \cite{StAh94}. The tricritical instability of the process, in particular, gives us a handle on the statistics of collapsing RBPs. We discuss the relation of our model to the asymmetric
Potts model and carefully analyze its symmetries. In the swollen phase, the
model has a high super-symmetry including translation and rotation invariance
in super-space and leads to the well known Parisi--Sourlas dimensional
reduction \cite{PaSo81}. At the collapse transition, super-rotation symmetry
is lost, and only translation invariance in superspace, i.e.,
Becchi--Rouet--Stora (BRS) symmetry~\cite{BRS75}, is restituted at the fixed point of the renormalization group. The appearance of BRS-symmetry shows that only tree-like RBPs \cite{Sha83,Ca2003} are relevant also at the collapse transition. We perform a 2-loop
renormalization group (RG) calculation, that corrects and extends the long
standing LI results for the collapse transition. Furthermore, we show that the
$\theta^{\prime}$-transition is characterized by a runaway of the RG flow
which suggests that this transition is a fluctuation induced first order
transition contrary to what has been assumed in recent numerical
studies \cite{HsGr05,HsNaGr05,DeHe83,HeSe96,SeVa94,FGSW92/94,JvR97/99/00}.

The outline of our paper is as follows: In Sec.~\ref{sec:modelling}, we derive
our dynamical field theoretical model starting from the Langevin equation for
a generalization of the so-called general epidemic process (GEP). We discuss
different limits of this model and recast it into different forms to reveal
the symmetry contents and to establish the connections to previous work in
particular that of LI and PS. In Sec.~\ref{sec:renColTran} we present the core
of our RG analysis with focus on the $\theta$-transition. We define our RG
scheme and we set up RG equations. We analyze the RG flow and its fixed
points, and we point out the implication of this flow for the $\theta^{\prime
}$-transition. In Sec.~\ref{sec:RBPobservables}, we extract from our RG results for various observables common in polymer physics.
In particular, we calculate scaling forms and critical exponents for the
$\theta$-transition.We also present results for the fractal dimension of the minimal path on RBPs at the collapse transition and in the swollen phase. At the collapse transition, in particular, this fractal dimension determines several other fractal properties since large RBPs are effectively tree-like. In Sec.~\ref{sec:concludingRemarks}, we give a few concluding remarks. There are three appendixes that present some additional information and some of the more technical aspects of our study.

\section{Modelling randomly branched polymers}
\label{sec:modelling}

In this section we develop our model for RBPs based on the GEP which is perhaps the most widely studied reaction diffusion
process in the universality class of dynamical isotropic percolation. To be
more specific, we use a generalization of this process that allows for a
tricritical instability. We will start out with the Langevin equation for this
generalized GEP which we will refine into a minimal model in the sense of
renormalized field theory. For background on field theory methods in general,
we refer to~\cite{Am84,ZJ02}. For background on dynamical field theory in the
context of percolation problems, we refer to~\cite{JaTa05}. For a related
approach to the somewhat simpler problem of directed randomly branched
polymers, see~\cite{JWS09}.

\subsection{Lattice animals}

Usually, one models RBPs by means of so-called lattice animals which are nothing but
clusters of connected sites on a regular lattice. One considers as the primary
quantity the number $\mathcal{A}(N,l,c)$ of all different configurations (up to translations) of a
single cluster (animal) which is a collection of $N$
sites, connected by $b\geq N-1$ bonds, $l$ cycles of the
bonds, and $c$ contacts (nearest-neighbor pairs of non-bonded sites).
The number of occupied bonds is then given by $b=l+N-1$. There is no
need for introducing a separate number $s$ of nearest-neighbor pairs
of occupied and non-occupied sites. This number is given by the relation
$\mathcal{N}N=2b+2c+s$, where $\mathcal{N}$ is the lattice
coordination number, which is equal to $2d$ on a simple hypercubic lattice.  The weighted animal number
\begin{equation}
\mathcal{A}_{N}(z_{cy},z_{co})=\sum_{l,c}\mathcal{A}(N,l,c)z_{cy}^{l}%
z_{co}^{c} \label{gew.An}%
\end{equation}
represents a general partition sum for the system. If one sets $z_{cy}$ to zero, the sum only includes tree configurations. It is well known that this partition function, also known as the generating function of lattice animals, can be obtained from the asymmetric $(n+1)$-state Potts model in the
limit $n\rightarrow0$ \cite{LuIs78,HaLu81,Con83}, and it is this connection, that stands behind the seminal earlier results on RBPs, cf.\ Sec.~\ref{sec:introduction}.

Typically, one considers the partition sum for large animals: $N\gg1$.
The phase diagram in this limit in terms of the fugacities $z_{cy}$ and
$z_{co}$ is shown in Fig.~\ref{fig:phaseDia}. The special curve
$z_{cy}=(z_{co}-1)z_{co}$, parametrized by a bond-probability $p$
as $z_{cy}=p/(1-p)^{2}$, $z_{co}=1/(1-p)$ defines a
bond-percolation model with the percolation probability $p=p_{c}$
depending on the specific type of the lattice. In general, if $N\gg1$,
there is a swollen phase for small fugacities, and a compact phase
separated by the collapse transition line $z_{co}(z_{cy})$ which
consists of two parts separated by the percolation point as a higher order
critical point. Whereas in the swollen phase $\mathcal{A}_{N}(z_{cy}%
,z_{co})\sim\kappa_{sw}(z_{cy},z_{co})^{N}N^{-\dot{\theta}}$ with
universal $\dot{\theta}$ and non-universal $\kappa_{sw}%
(z_{cy},z_{co})$, one finds at least for the left part of the
transition line the scaling law
\begin{equation}
\mathcal{A}_{N}(z_{cy},z_{co}(z_{cy}))\sim\kappa(z_{cy})^{N}N^{-\theta}
\label{as.gew.An}%
\end{equation}
with non-universal $\kappa(z_{cy})$, and universal $\theta$
in general different from $\dot{\theta}$ \cite{HsGr05}. The percolation point as a separating point on the transition line with higher order critical behavior has a $\theta_{perc}$ which is
in general different from $\theta$ and $\dot{\theta}$. Only in mean-field theory
(Landau approximation) these exponents are equal: $\theta=\dot{\theta}=\theta_{perc}=5/2$.

Other fundamental quantities are given by correlations of sites on
the cluster. The correlation function may be defined by
\begin{equation}
G_{N}(\mathbf{r},\mathbf{r}^{\prime})=\frac{1}{\mathcal{A}_{N}(z_{cy},z_{co}%
)}\sum_{l,c}\mathcal{A}(N,l,c;\mathbf{r},\mathbf{r}^{\prime})z_{cy}^{l}%
z_{co}^{c}\,, \label{An-Korr}%
\end{equation}
where $\mathcal{A}(N,l,c;\mathbf{r},\mathbf{r}^{\prime})$ is
the total number of clusters with $N$ sites, $l$ loops, and
$c$ contacts, containing the lattice sites $\mathbf{r}$ and
$\mathbf{r}^{\prime}$. Of course it is
\begin{equation}
\sum_{\mathbf{r}}\mathcal{A}(N,l,c;\mathbf{r},\mathbf{r}^{\prime})=N\mathcal{A}(N,l,c)
\end{equation}
The radius of gyration $R_{N}$ is then defined by
\begin{equation}
R_{N}^{2}=\frac{1}{2dN} \sum_{\mathbf{r},\mathbf{r}^{\prime}}(\mathbf{r}%
-\mathbf{r}^{\prime})^{2}G_{N}(\mathbf{r},\mathbf{r}^{\prime}) \, .
\end{equation}
For $N\gg1$, it shows also an universal scaling law
\begin{equation}
R_{N}\sim N^{\nu_{A}}\,. \label{Gyr.Rad.}%
\end{equation}
The fractal dimension $d_{f}=1/\nu_{A}$ is different at the
transition line from its value in the swollen phase and at the separating
percolation point. However, in mean-field theory it has the uniform value
$d_{f}=4$. Of course, in the compact phase, the fractal dimension is
always equal to the lattice dimension $d$.

\subsection{Reactions, Langevin equation, and dynamic response functional}

The model that we are about to develop is in the spirit of Landau's ideas for modeling second-order phase transitions, i.e., it is a mesoscopic model that focuses on general principles unifying
processes belonging to the same universality class and is therefore
necessarily phenomenological \cite{JaTa05}. To set the stage, however, we find it worthwhile to discuss in some detail a specific model belonging to the RBP universality class, viz.\ a generalization of the GEP. The reaction-diffusion equations defining this process will nurture our intuition and will help us to establish our ideas.

The following  generalization of the GEP is a variant
of a process that we have introduced for the description of tricritical
isotropic percolation \cite{JaMuSt04}. We denote by $X(\mathbf{r})$ an agent, i.e., an infected individual, at site
$\mathbf{r}$. An agent can infect a neighboring site $\mathbf{r+\delta}$ via the percolation step
\begin{equation}
X(\mathbf{r})\rightarrow X(\mathbf{r})+X(\mathbf{r+\delta})\,.\label{Perc}%
\end{equation}
This fundamental reaction gives rise to spreading and branching of the
epidemic. The agents can spontaneously become immune (or decay) and produce spam as a marker of the agent through the reactions
\begin{subequations}
\begin{align}
X(\mathbf{r}) &  \rightarrow Z(\mathbf{r})\,,\label{dec}
\\
X(\mathbf{r}) &  \rightarrow X(\mathbf{r})+Z(\mathbf{r})\,,\label{spam}%
\end{align}
\end{subequations}
where $Z(\mathbf{r})$ denotes an immune individual or spam at site $\mathbf{r}$. In the language of forest fires, the $Z(\mathbf{r})$ are also referred to as debris. It is the debris left behind by the epidemic that forms the clusters which serve us as prototypes for RBPs. Their self-avoidance or excluded volume interaction is modelled with help of the reaction
\begin{equation}
X(\mathbf{r})+kZ(\mathbf{r})\rightarrow(k+1)Z(\mathbf{r})\,,\label{suppr}%
\end{equation}
where $k=1,2,\cdots$, which dampens the epidemic. A mechanism for the RBPs to compactify is introduced into the process through the reaction
\begin{equation}
X(\mathbf{r-\delta})+Z(\mathbf{r+\delta})\rightarrow X(\mathbf{r-\delta
})+X(\mathbf{r})+Z(\mathbf{r+\delta})\,,\label{attrac}%
\end{equation}
which simulates an effective attraction of the agents by the debris.

Having these reactions, one possible way to proceed would be to reformulate the corresponding master-equation in terms of bosonic creation and annihilation operators and then to produce a field theoretic action from these operators via coherent state path integrals \cite{TaHoVo05}. However, we prefer to extract directly the mesoscopic Langevin equations that incorporate the universal features of the above reactions, namely the percolation of agents, their spontaneous decay, their suppression and possible effective attraction by the debris, and the possible existence of vacua without agents as absorbing states of the system.

The primary density-fields describing our generalized GEP are the field of
agents $n(\mathbf{r},t)$ and the field of the inactive debris $m(\mathbf{r}%
,t)=\lambda\int_{-\infty}^{t}dt^{\prime}\,n(\mathbf{r},t^{\prime})$ which
ultimately forms the polymer cluster. A non-Markovian Langevin equation
describing such a process, and represents therefore the universality class, is
given by
\begin{equation}
\lambda^{-1}\partial_{t}n=\nabla^{2}n+c\nabla m\cdot\nabla n-\Big[r+g^{\prime
}m+\frac{f^{\prime}}{2}m^{2}\Big]n+\zeta\,. \label{Lang-Eq}%
\end{equation}
Here, the parameter $r$ tunes the "distance" to the percolation threshold.
Below this threshold, i.e., in the absorbing phase, $r$ is positive. Throughout this paper, we will assume that the system is deep in the absorbing phase. In this case, a typical final cluster generated from an additional source $q\delta
(\mathbf{r})\delta(t)$ of agents adding such a source is equivalent to specifying an initial condition for the process) consists of $N=\langle\int d^{d}r\,m(\mathbf{r},\infty)\rangle\approx q/r$ debris-particles, and has a mean diameter $1/\sqrt{r}$. However, we are
interested in the large non-typical clusters, the rare events of the
stochastic process, with $N\gg q/r$. We know from percolation theory
\cite{StAh94} that these clusters belong to the universality class of
lattice animals. Hence, they are the same in a statistical sense as randomly branched polymers as far as their universal properties go. The
gradient-term proportional to $c$ describes the attractive influence of the debris on the agents if $c$ is
negative (as a negative contribution to $g'$ does). At this point other forms of gradient-terms like
$m\nabla^{2}n$ and $n\nabla^{2}m$ are conceivable. However al long as we include any one of these gradient terms into our theory, an omission of the other gradient terms has no effect on the final results, and we choose to work with the term proportional to $c$ only for simplicity. For usual percolation problems (ordinary or
tricritical), these gradient  terms are irrelevant. As long as $g^{\prime}>0$, the
second order term $f^{\prime}m^{2}$ is irrelevant near the transition point
and the process models ordinary percolation near $r=0$ \cite{Ja85} or non-typical very large clusters, the swollen RBPs, for $r>0$. We permit both signs of
$g^{\prime}$ (negative values of $g^{\prime}$ correspond to an attraction of the agents by the debris, see above). Hence, our model allows
for a tricritical instability (tricritical percolation near $r=0$ \cite{JaMuSt04}
or the collapse transition of the RBPs for $r>0$).
Consequently we need the second order term $f^{\prime}>0$ (which represents the self-avoidance property)  to limit the density to finite values. Physically it
originates from the suppression of agents by the debris. The Gaussian
noise-source $\zeta(\mathbf{r},t)$ has correlations
\begin{align}
\overline{\zeta(\mathbf{r},t)\zeta(\mathbf{r}^{\prime},t^{\prime})}  &
=\Big[\lambda^{-1}gn(\mathbf{r},t)\delta(t-t^{\prime})-fn(\mathbf{r}%
,t)n(\mathbf{r^{\prime}},t^{\prime})\Big]\nonumber\\
&  \qquad\qquad\qquad\times\delta(\mathbf{r}-\mathbf{r}^{\prime})\,.
\label{Abs-noise}%
\end{align}
The process is assumed to be locally absorbing, and thus all terms in the
noise-correlation function contain at least one power of $n$. The first part
of the noise correlation takes into account that the agents decay
spontaneously, and thus $g>0$. The non-Markovian term proportional to $f$ simulates the
anticorrelating or, respectively, correlating (from attraction) behavior of the
noise in regions where debris has already been produced with $f$ being negative if the attraction effects are overwhelming.

Two points are worth mentioning at this stage: (i) For the Langevin-equation with the local noise to be meaningful mathematically, an appropriate cut-off procedure of long wavelengths has to be used.
(ii) The stochastic process (\ref{Lang-Eq}) with $c=r=g^{\prime}=f^{\prime}=g=0$ but $f>0$ belongs to the
universality class of self-avoiding random walks (SAW), and generates
therefore the statistics of linear polymers \cite{Sch99}.

To proceed towards a field theoretic model, the Langevin equations are now
transformed into a stochastic response functional in the Ito-sense
\cite{Ja76,DeDo76,Ja92,JaTa05}%
\begin{align}
\mathcal{J}  &  =\int d^{d}x\Big\{\lambda\int dt\tilde{n}\Big[\lambda
^{-1}\partial_{t}-\nabla^{2}-c\nabla m\cdot\nabla+r+g^{\prime}m\nonumber\\
&\qquad  +\frac{f^{\prime}}{2}m^{2}-\frac{g}{2}\tilde{n}\Big]n+\frac{f}%
{2}\Big[\lambda\int dt\,\tilde{n}n\Big]^{2}\Big\}\ . \label{StochFu}%
\end{align}
With this functional, we now have a vantage point for the calculation of statistical quantities via path-integrals with the exponential weight $exp(-\mathcal{J})$. When a source-term  $(\tilde{h},\tilde{n})$ is added, where $\tilde{h}(\mathbf{r},t)=\tilde{h}_0(\mathbf{r},t)=q\delta(\mathbf{r})\delta(t)$ and $(..,..)$ denotes an integral of a product of two fields over space and time, this functional describes, in particular, the statistics of clusters of debris generated by the stochastic process (\ref{Lang-Eq}) from a source of $q$ agents at the point $\mathbf{r}=0$ at time
zero. Denoting by $\operatorname*{Tr}\bigl[\ldots\bigr]$ the
functional integration over the fields, we generally have
\begin{equation}
\operatorname*{Tr}\bigl[\exp\bigl(-\mathcal{J}+(\tilde{h},\tilde{n})+(h,n)\bigr]=1
\end{equation}
if $h$ \underline{or} $\tilde{h}$ are zero. The first property
follows from causality whereas the second one originates from the absorptive
properties of the process. Note that the role of causality and adsorptivity
can be interchanged via the duality transformation $m(\mathbf{r}%
,t)\longleftrightarrow-\tilde{n}(\mathbf{r},-t)$ \cite{Ja85,Ja05,JaTa05}.

\subsection{Branched polymers as rare events}

Averaging an observable $\mathcal{O}[n]$ over final clusters
of debris (the RBPs) of a given mass $N$ generated from a source
$\tilde{h}(\mathbf{r},t)=q\delta(\mathbf{r})\delta(t)$ of agents at the origin
$\mathbf{r}=0$ at time $t=0$ leads to the quantity \cite{Ja85,JaTa05,Ja05}%
\begin{align}
&  \langle\mathcal{O}\rangle_{N}\mathcal{P}(N)=\left\langle\mathcal{O}[n]
\delta(N-\mathcal{M})\exp\bigl((\tilde{h},\tilde{n}%
)\bigr)\right\rangle \nonumber\\
&  =\operatorname*{Tr}\Big[\mathcal{O}[n]\delta(N-\mathcal{M})\exp
\bigl(-\mathcal{J}+q\tilde{n}(0,0)\bigr)\Big]\nonumber\\
&  \simeq q\operatorname*{Tr}\Big[\mathcal{O}[n]\tilde{n}(0,0)
\delta(N-\mathcal{M})\exp\bigl(-\mathcal{J}\bigr)\Big]\,, \label{Erw-Wert}%
\end{align}
where
\begin{equation}
\mathcal{P}(N)=\langle\delta(N-\mathcal{M})\exp\bigl(q\tilde{n}%
(0,0)\bigr)\rangle\label{defP}%
\end{equation}
is the probability distribution for finding a cluster of mass $N$.
\begin{equation}
\mathcal{M}=\int d^{d}rdt\,\lambda n(\mathbf{r},t)=\int d^{d}r\,m_{\infty
}(\mathbf{r})\, \label{Masse}%
\end{equation}
is the total mass of the debris. The field $m_{\infty
}(\mathbf{r})=m(\mathbf{r},t=\infty)$ describes the distribution of
the debris after the epidemic has become extinct. Since the probability
distribution should be proportional to the number of different configurations,
we expect by virtue of universality arguments the following proportionality
between the probability distribution $\mathcal{P}(N)$ and the
lattice animal number $\mathcal{A}_{N}$ for asymptotically large $N$:
\begin{equation}
\mathcal{A}_{N}\sim N^{-1}\kappa_{0}^{N}\mathcal{P}(N)\,, \label{A_zu_P}%
\end{equation}
where $\kappa_{0}$ is an effective coordination number of the underlying
lattice. The fugacities in $\mathcal{A}_{N}(z_{cy},z_{co})$
are then considered as analytical functions of the different parameters
in the response functional $\mathcal{J}$ or vice versa. The factor $N^{-1}$
arises in Eq.~(\ref{A_zu_P}) because the generated clusters are
rooted at the source at the point $\mathbf{r}=0$, and each site of a
given lattice animal may be the root of given cluster. Hence, we
expect a scaling
\begin{equation}
\mathcal{P}(N)\sim N^{1-\theta}p_{0}^{N} \label{PzuA}
\end{equation}
with an universal scaling exponent $\theta$ but non-universal $p_{0}$.

In actual calculations, the delta function appearing in averages like in
Eq.~(\ref{defP}) is hard to handle. This problem can be simplified by using
Laplace-transformed observables like, e.g., the Laplace transformation of $\mathcal{P}(N)$, which are functions of a variable
conjugate to $N$, say $z$,
\begin{equation}
\mathcal{P}(N)=\int_{\sigma-i\infty}^{\sigma+i\infty}\frac{dz}{2\pi
i}\,\mathrm{e}^{zN}\,\langle\exp\bigl(-z\mathcal{M+}q\tilde{n}%
(0,0)\bigr)\rangle\,, \label{Inv-Lapl}%
\end{equation}
and applying inverse Laplace transformation (where all the
singularities of the integrand lie to the left of the integration path) in the end. Note that the relationship between $\mathcal{P}(N)$ and $\mathcal{A}_{N}$ signals the existence of a singularity $\sim (z-z_c)^{\theta -2}$ of the integrand in Eq.~(\ref{Inv-Lapl}) at some critical value $z_c$. The switch to Laplace-transformed observables can be done in a pragmatic way by
augmenting the original $\mathcal{J}$ with a term $z\mathcal{M}$ and then
working with the new response functional
\begin{equation}
\mathcal{J}_{z}=\mathcal{J}+z\mathcal{M}\,. \label{Def_Jz}%
\end{equation}
Denoting averages with respect to the new functional by $\langle\ldots
\rangle_{z}$, and defining
\begin{equation}
q\Phi(z)=\ln\langle\exp(q\tilde{n})\rangle
_{z}\approx q\langle\tilde{n}\rangle_{z}
\end{equation}
for small $q$, we get by using Jordans lemma that the
asymptotic behavior for large $N$ is given by
\begin{align}
\mathcal{P}(N)  &  =\int_{\sigma-i\infty}^{\sigma+i\infty}\frac{dz}{2\pi
i}\,\exp\bigl[zN+q\Phi(z)\bigr]\nonumber\\
&  =\mathrm{e}^{z_{c}N+q\Phi(z_{c})}\int\frac{dz^{\prime}}{2\pi i}%
\,\exp\bigl[z^{\prime}N\nonumber\\
&  +q(\Phi(z_{c}+z^{\prime})-\Phi(z_{c}))+O(q^{2})\bigr]\nonumber\\
&  \approx q\mathrm{e}^{z_{c}N+q\Phi(z_{c})}\int_{0}^{\infty}dx\,\frac
{\operatorname*{Disc}\Phi(z_{c}-x)}{2\pi i}\mathrm{e}^{-xN}\,,
\label{Probab_N}%
\end{align}
where the last row gives the asymptotics for large $N$ and small $q$.
Here, $z_{c}$ is the first singularity of $\Phi(z)$, which
as we will show is a branch point on the negative real axis, and the
contour of the path integral is deformed into a path above and below
the branch cut beginning at the singularity. $\operatorname*{Disc}\Phi$
denotes the discontinuity of the function $\Phi$ at the branch cut. The
non-universal factor $q\mathrm{e}^{z_{c}N+q\Phi(z_{c})}$ depending
exponentially on $N$ is common to all averages defined by
Eq.~(\ref{Erw-Wert}) and therefore cancels from all mean values
$\langle\mathcal{O}\rangle_{N}$.

\subsection{Mean-field theory}

Before we assent to the heights of field theory (or decent to its depths, if the reader prefers,
we first apply a mean-field approximation to our
theory, i.e., we solve the functional integrals with
the weight $\exp(-\mathcal{J}_{z})$ using a saddle-point approximation. The linear term in
$\mathcal{J}_{z}$  that is proportional to the Laplace-variable $z$
leads to a non-zero saddle-point value of the field $\tilde{n}$:%
\begin{equation}
\tilde{n}_{SP}=\Phi(z)\,. \label{mf-OrdPar}%
\end{equation}
Therefore, shifting this field, $\tilde{n}\rightarrow\tilde{n}+\Phi$, so that
\begin{equation}
\langle\tilde{n}\rangle_{z}:=\operatorname*{Tr}\bigl[\tilde{n}\exp
(-\mathcal{J}_{z})\bigr]=0 \label{TadBed}%
\end{equation}
the harmonic (Gaussian) part of $\mathcal{J}_{z}$ becomes
\begin{align}
\mathcal{J}_{z}^{(0)}  &  =\int d^{d}x\Big\{\lambda\int dt\tilde
{n}\Big[\lambda^{-1}\partial_{t}-\nabla^{2}+(r-g\Phi)\Big]n\nonumber\\
& \qquad +\frac{c\Phi}{2}(\nabla m_{\infty})^{2}+\frac{\Phi}{2}(g^{\prime}%
+f\Phi)m_{\infty}^{2}\nonumber\\
& \qquad +(z+r\Phi-\frac{g}{2}\Phi^{2})m_{\infty}\Big\}\,. \label{J-harm}%
\end{align}
Here, we have implied that the saddlepoint-value of $m_{\infty}$
is zero, i.e., we have assume that $\rho=(g^{\prime
}+f\Phi)\Phi$ is positive. If $\rho=0$, which is the case
near the tricritical instability of our stochastic process, a phase transition
to a positive value of $\langle m_{\infty}\rangle$ sets in. Whether or not this transition is the anticipated collapse transition deserves some further scrutiny. A shift
\begin{align}
\label{shiftTrafo}
\tilde{n}\rightarrow\tilde{n}+\alpha
m_{\infty}
\end{align}
 (which does not change the condition (\ref{TadBed}))
changes $\Phi(g^{\prime}+f\Phi)$ to $\Phi(g^{\prime}+f\Phi
)+\alpha\tau$, where $\tau=r-g\Phi$. The special value
$\alpha=-c\Phi/2$ eliminates the gradient-term $\sim(\nabla
m_{\infty})^{2}$ and hence $\rho=0$
signals the collapse only if $\tau$ goes to zero which is
indeed the critical value corresponding to large clusters with $N\gg1$.
This can be seen from the saddlepoint condition $h=z+r\Phi-g\Phi
^{2}/2=0$ that leads to
\begin{equation}
g\Phi(z)=r-\sqrt{r^{2}+2gz}\,. \label{mf-Phi}%
\end{equation}
Thus, the meanfield-solution shows a branch-point singularity at
$z_{c}=-r^{2}/2g$, and $\tau(z)=\sqrt{r^{2}+2gz}$ becomes
zero at this singularity.

Until now, we have kept the gradient term proportional to $c$ in our theory. The discussion in the last paragraph revealed that this term is redundant in the sense of field theory as it can be eliminated via the shift transformation (\ref{shiftTrafo}). Hence, we will formally set $c=0$ unless noted otherwise.

Next, let us calculate $\mathcal{P}(N)$ from Eq.~(\ref{Probab_N}).
Inserting $\Phi(z)$ from Eq.~(\ref{mf-Phi}), we easily
obtain the probability density of branched polymers with size $N$
in mean-field approximation,
\begin{equation}
\mathcal{P}(N)=\frac{q}{\sqrt{2\pi g}}N^{-3/2}\exp\Big(\frac{rq}{g}%
-\frac{r^{2}}{2g}N-\frac{q^{2}}{2g}N^{-1}\Big)\,. \label{mf-Prob}%
\end{equation}
The maximum of this distribution is found at $N=N_{0}=q/r$.
For $N\gg q/r$, the distribution drops down exponentially.
However, this is the region of rare events of our stochastic process where the
large branched polymers are found. Hence, small $q$ means effectively
$q\ll rN$, and $q=1$ is `small' in this region. Combining
 Eqs.~(\ref{as.gew.An}) and (\ref{A_zu_P}), we obtain the asymptotic
result
\begin{equation}
\mathcal{P}(N)\sim N^{-3/2}\exp(-r^{2}N/2g) \, ,
\end{equation}
and the well-known mean-field animal exponent $\theta=5/2$ common to the
swollen phase, the percolation point, as well as the collapse
transition-line.

Now, we calculate the monomer distribution (the distribution of the
debris-particles) of a single large cluster rooted at the point
$\mathbf{r}=0$. We recall from our remarks above that such a root is
represented field theoretically by an insertion of the field $\tilde{n}(0,0)$. According
to Eq.~(\ref{Erw-Wert}), the monomer distribution is given by the
inverse Laplace transformation of the correlation function calculated with the
harmonic response functional (\ref{J-harm}):
\begin{align}
&  \langle m_{\infty}(\mathbf{r})\tilde{n}(0,0)\rangle_{z} =G_{1,1}%
(\mathbf{r};z)=\int_{\mathbf{k}}\frac{\exp(i\mathbf{k}\cdot\mathbf{r)}}%
{\tau(z)+\mathbf{k}^{2}}\nonumber\\
&  \qquad=\int_{0}^{\infty}\frac{ds}{(4\pi s)^{d/2}}\exp\bigl(-s\tau
(z)-\mathbf{r}^{2}/4s\bigr)\,.
\end{align}
It follows that
\begin{align}
G_{N}(\mathbf{r})  &  =\frac{1}{\mathcal{P}(N)}\int_{\sigma-i\infty}%
^{\sigma+i\infty}\frac{dz}{2\pi i}\,\mathrm{e}^{zN}G_{1,1}(\mathbf{r}%
;z)\nonumber\\
&  =\frac{g}{(4\pi)^{d/2}}\int_{0}^{\infty}\frac{ds}{s^{d/2-1}}\exp
\bigl(-gs^{2}/2N-\mathbf{r}^{2}/4s\bigr)\,. \label{mf-KorrF}%
\end{align}
This function can be written in terms of generalized hypergeometric
series $_{0}F_{2}$, however, we prefer the integral representation shown in Eq.~(\ref{mf-KorrF}).
Easily, we verify the sum rule
\begin{equation}
\int d^{d}rG_{N}(\mathbf{r})=N\,.
\end{equation}
The radius of gyration $R_{N}$ can be calculated straightforwardly from its
definition,
\begin{equation}
R_{N}^{2}=\frac{1}{Nd}\int d^{d}rG_{N}(\mathbf{r})\mathbf{r}^{2}=\bigl(2\pi
N/g\bigr)^{1/2}\,. \label{mf-GyrRad}%
\end{equation}
Hence, the gyration exponent is $\nu_{A}=1/4$ as anticipated.
The integral representation (\ref{mf-KorrF}) yields the asymptotic forms of
the monomer distribution for $\left\vert \mathbf{r}\right\vert \ll R_{N}$
\begin{equation}
G_{N}(\mathbf{r})\sim\frac{1}{\left\vert \mathbf{r}\right\vert ^{d-4}}\,,
\end{equation}
and
\begin{equation}
G_{N}(\mathbf{r})\sim\frac{N}{R_{N}^{d}}\left(  \frac{R_{N}}{\left\vert
\mathbf{r}\right\vert }\right)  ^{(d-2)/3}\exp\left(  -\frac{3\pi^{1/3}}%
{4}\left(  \frac{\left\vert \mathbf{r}\right\vert }{R_{N}}\right)
^{4/3}\right)
\end{equation}
if $\left\vert \mathbf{r}\right\vert \gg R_{N}$. We see that
the monomer-distribution in the fractal interior of the cluster has a fractal
dimension $d_{f}=4$ independent of $N$. The distribution in
the outer region drops down exponentially in $\left\vert \mathbf{r}%
\right\vert $, however, with an exponent $4/3=1/(1-\nu_{A})$.
Besides the exponential factor, the distribution decreases
algebraically with an exponent $(d-2)/3=(d/2-d\nu_{A}+2-\theta)/(1-\nu_{A})$.
We will show later on that these scaling relations comprising the independent critical exponents $\theta$ and $\nu_A$ hold generally and are not restricted to the mean-field approximation.

Another interesting quantity is the correlation of two roots. Evidently, two roots can either belong to one cluster or they can belong to two separate clusters. Their correlation function is of some value in polymer physics because it determines the second virial coefficient of the equation of state of a dilute solution of branched polymers. The connected part of this correlation function, i.e., the cumulant, is given by
\begin{align}
&  \langle\tilde{n}(\mathbf{r},0)\tilde{n}(0,0)\rangle_{z}^{(cum)}
=C(\mathbf{r};z)=-\rho\int_{\mathbf{k}}\frac{\exp(i\mathbf{k}\cdot\mathbf{r)}%
}{\bigl(\tau(z)+\mathbf{k}^{2}\bigr)^{2}}\nonumber\\
&  \qquad=-\rho\int_{0}^{\infty}\frac{ds}{(4\pi s)^{d/2}}s\exp\bigl(-s\tau
(z)-\mathbf{r}^{2}/4s\bigr)\,.
\end{align}
Inverse Laplace-transformation leads to
\begin{align}
C_{N}(\mathbf{r})&\sim\frac{-\rho N^{3/2}}{R_{N}^{d}}\left(  \frac{R_{N}%
}{\left\vert \mathbf{r}\right\vert }\right)  ^{(d-4)/3}
\nonumber \\
& \times \exp\left[  -\frac
{3\pi^{1/3}}{4}\left(  \frac{\left\vert \mathbf{r}\right\vert }{R_{N}}\right]
^{4/3}\right)
\end{align}
in the region $\left\vert \mathbf{r}\right\vert \gg R_{N}$, where $N$ should be understood here as the total number of
monomers. Since the correlation of roots on the same cluster goes down
proportionally to the density of monomers on one single cluster, the increasing
behavior of the fraction
\begin{equation}
C_{N}(\mathbf{r})/G_{N}(\mathbf{r})\sim-\rho N^{1/2}\left(
\frac{\left\vert \mathbf{r}\right\vert }{R_{N}}\right)  ^{2/3}%
\end{equation}
results mainly from the interaction of two separate clusters. They are
repelling one another if $\rho$ is positive, and attracting one another for negative $\rho$.
The sharp difference between repelling and attracting is a clear
signature of the collapse transition located at $\rho=0$. Note that a
contribution to $\rho$ proportional to $\tau$ as discussed
above leads only to a change of the pivotal factor $\rho N^{1/2}$ of
order $1$ since $\tau(z)$ converts to a term $\sim
N^{-1/2}$ through the inverse Laplace-transformation.

\subsection{Dynamical response functional revisited}

Now, we return to our response functional to refine it into a form that suits us best for our actual field theoretic analysis. As discussed above, the gradient term proportional to $c$ is redundant. To eliminate this term, we apply to the  field $\tilde{n}$ the shift and mixing transformation
\begin{equation}
\tilde{n}(\mathbf{r},t)\rightarrow\tilde{n}(\mathbf{r},t)+\Phi-c\Phi
m_{\infty}(\mathbf{r})\,., \label{RedTraf}%
\end{equation}
where $\Phi$ is a free parameter at this stage. Defining in consistency with our mean-field considerations above $\tau=r-g\Phi$, $\rho=(g^{\prime}+f\Phi
)\Phi-c\Phi\tau$, $h=z+r\Phi-g\Phi^{2}/2$, the stochastic functional
$\mathcal{J}_{z}$ (\ref{Def_Jz}) takes the form
\begin{align}
\mathcal{J}_{z}  &  =\int d^{d}x\Big\{\lambda\int dt\,\tilde{n}\Big[\lambda
^{-1}\partial_{t}+\tau-\nabla^{2}+g_{2}^{\prime}m-\frac{g_{2}}{2}\tilde
{n}\nonumber\\
&  +g_{1}m_{\infty}\Big]n+\Big[\frac{\rho}{2}m_{\infty}^{2}+\frac{g_{0}}%
{6}m_{\infty}^{3}+hm_{\infty}\Big]\Big\}\,. \label{Jz}%
\end{align}
Here, we could have set $\tau$ equal to zero by exploiting that $\Phi$ is a free parameter. Instead of doing so, we rather keep $\tau$ in our theory as a small free parameter. We will see later on that keeping $\tau$ comes in handy for renormalization purposes.
In Eq.~(\ref{Jz}), we have eliminated couplings that are of more than third order
in the fields because they are irrelevant. We do not write down in
detail the relatively uninteresting relations between the new third-order
coupling constants and the old ones. Note that $\mathcal{J}_{z}$ contains two
similar couplings: $g_{2}^{\prime}\tilde{n}nm$ and $g_{1}\tilde{n}nm_{\infty}%
$. Whereas the first coupling respects causal ordering, which means that
$\tilde{n}$ is separated by an infinitesimal positive time-element from the
$nm$-part resulting from the Ito-calculus \cite{Ja92}, the second one respects
causality only between $\tilde{n}$ and $n$. In contrast to the $m$-part, the
$m_{\infty}$-part contains all the $n$ with times that lie in the past and in
the future of $\tilde{n}$. This property is the heritage of the
time-delocalized noise term. Even if we had disregarded the noise term proportional to $f$ in Eq.~(\ref{Abs-noise}) initially, the $\tilde{n}nm_{\infty}$-coupling would be generated by coarse graining, and hence it must be ultimatively incorporated into the theory to yield a renormalizable theory.

The relevance of the different terms in $\mathcal{J}_{z}$ follows from their
dimensions with respect to an inverse length scale $\mu$ such that time scales as $\mu^{-2}$. Fundamentally, one has to decide which
parameters are the critical control-parameters going to zero in mean-field
theory. As we have seen, at the collapse transition these are $\tau\sim
\rho\sim\mu^{2}$, and $h\sim\mu^{(d+2)/2}.$ The dimensions of the fields are
then given by $\tilde{n}\sim m\sim\mu^{(d-2)/2}$, and $n\sim\mu^{(d+2)/2}$. It
follows that all the coupling constant $g_{0}$, $g_{1}$, $g_{2}$, and
$g_{2}^{\prime}$ have the same dimension $\mu^{(6-d)/2}$. Note that $\tilde
{n}$ is tied always to at least one factor of $n$ as a result of absorptivity
of the process. Hence, we have retained all the couplings that are relevant
for $d\leq6$ spatial dimensions, and the model is renormalizable below the
upper critical dimension $d_{c}=6$ of the collapse transition. The situation is
different if $\rho$ is a finite positive quantity, that is in the swollen
phase. Then $\rho$ can be absorbed into the fields by a scale transformation which amounts to
formally setting $\rho=2$. The field dimensions then become $m\sim\mu^{d/2}$,
$n\sim\mu^{(d+4)/2}$, and $\tilde{n}\sim\mu^{(d-4)/2}$. It follows that $h\sim
\mu^{d/2}$, $g_{0}\sim\mu^{-d/2}$, $g_{1}\sim\mu^{(4-d)/2}$, $g_{2}^{\prime}\sim\mu^{(2-d)/2}$, and $g_{2}\sim\mu^{(8-d)/2}$. Hence, in the swollen
phase only $g_{2}=g$ is relevant, now below $8$ spatial dimensions. The other
couplings can be safely removed.

\subsection{Quasi-static limit and ghosts}

In the following, we focus on the static properties of the generated clusters
after the epidemic has become extinct. Here, we are interested only in
time-independent static expectation values of the form $\langle\prod
_{i}m_{\infty}(\mathbf{r}_{i})\prod_{j}\tilde{n}(\mathbf{r}_{j},0)\rangle$. Thus, we take the quasi-static limit \cite{Ja85,JaTa05,JaMuSt04,JaLy94}, see
Appendix~\ref{app:quasi-staticLimit}, by setting $\tilde{n}(\mathbf{r},t)\rightarrow\tilde{n}_{0}%
(\mathbf{r})=:\varphi(\mathbf{r})$ in the dynamic response functional
$\mathcal{J}_{z}$. We rename $m_{\infty}(\mathbf{r})=:\tilde{\varphi
}(\mathbf{r})$ and get%
\begin{align}
\mathcal{J}_{z}  &  \rightarrow\mathcal{H}_{qs}=\int d^{d}x\Big\{\tilde
{\varphi}\bigl(\tau-\nabla^{2}\bigr)\varphi+\frac{\rho}{2}\tilde{\varphi}%
^{2}+h\tilde{\varphi}\nonumber\\
&  +\frac{g_{0}}{6}\tilde{\varphi}^{3}+g_{1}\tilde{\varphi}{\varphi\cdot
}\tilde{\varphi}+\frac{1}{2}\tilde{\varphi}\bigl(g_{2}^{\prime}\tilde{\varphi
}-g_{2}\varphi\bigr)\varphi\Big\}\,, \label{H-quasi}%
\end{align}
where we have denoted the original time-delocalisation of the $\tilde
{n}nm_{\infty}$-term by a separating dot in $\tilde{\varphi}{\varphi\cdot
}\tilde{\varphi}$. Using the quasistatic limit, one has to be careful to
account for the former causal ordering of fields in the diagrammatic
perturbation expansion. This means that one has to rule out diagrams with
closed propagator-loops. But note that only the $\tilde{\varphi}{\varphi}%
$-part of the $\tilde{\varphi}{\varphi\cdot}\tilde{\varphi}$-term can
contribute to such a closed loop.

\begin{figure}[ptb]
\centering{\includegraphics[width=3cm]{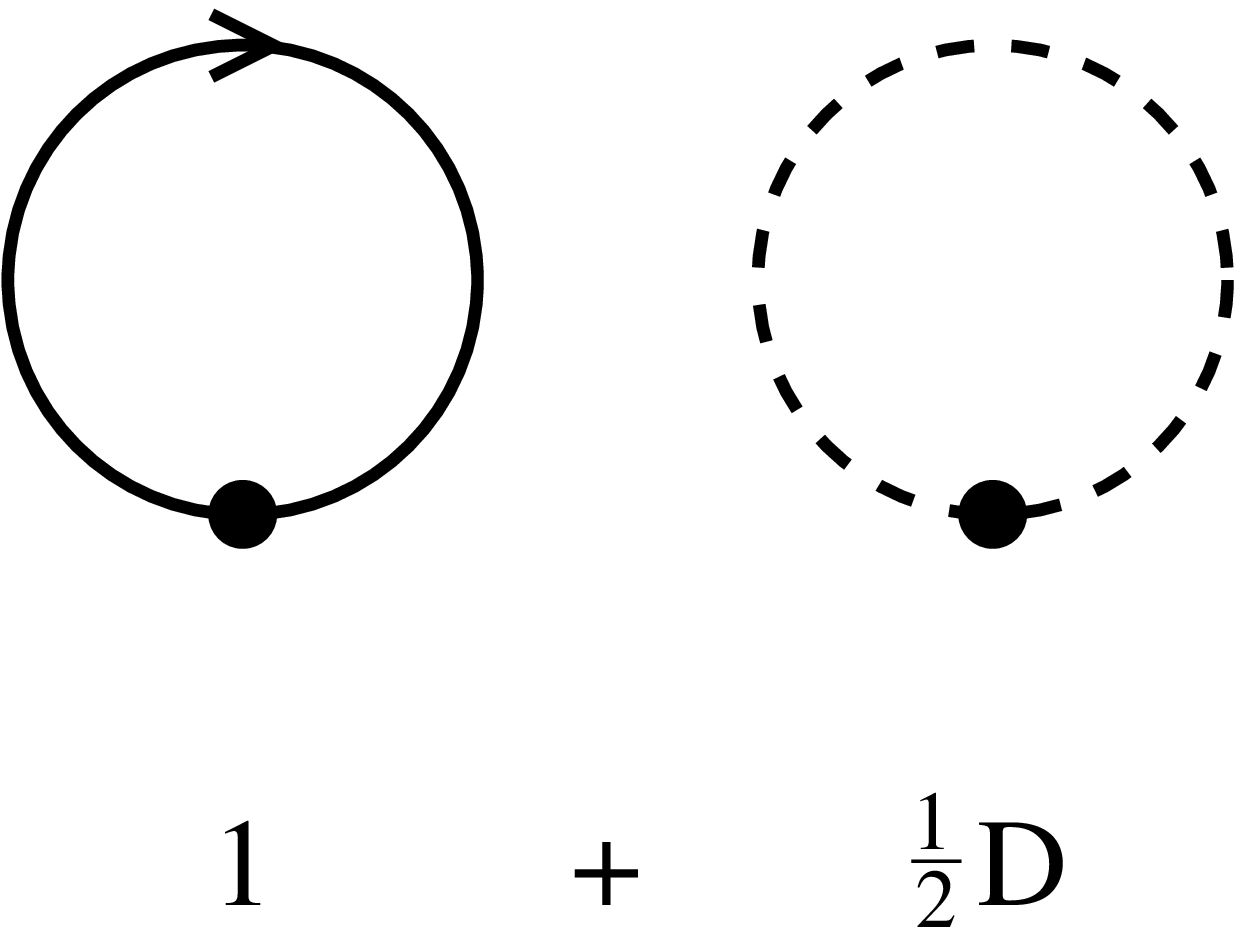}} \caption{1-loop non-causal
diagrams and their cancellation by $D=-2$ bosonic ghost fields.}%
\label{fig:cancellation1Loop}%
\end{figure}

Of course these additional rules make the perturbation expansion very clumsy
in higher loop-order calculations. Fortunately there exists an elegant way to
overcome these difficulties associated with the additional rules by introducing
so-called ghost fields whose sole purpose is to generate additional diagrams
that cancel any diagrams with non-causal propagator-loops. Such a procedure
does not change the physical content of the theory but simplifies calculations
and makes it easier to find higher symmetries. To one-loop order, non-causal
loops are easily cancelled by a corresponding loop of contrary sign. The ghost
fields for producing such loops that come to mind first are a pair of fermionic
fields. Note, however, that $D$ independent similar bosons can also create a
loop with a negative sign in the limit $D\rightarrow-2$, see
Fig.~\ref{fig:cancellation1Loop}.

To be more specific, the ghost fields that we use are $D$ independent bosonic fields $(\psi_{1},\ldots,\psi_{D})$,
in the limit $D\rightarrow-2$ which is taken at the end of the calculation. These ghosts are incorporated into our theory be adding the term
\begin{equation}
\frac{1}{2}\sum_{k=1}^{D}\psi_{k}\Big[\tau_{0}-\nabla^{2}+\bigl((g_{1}%
+g_{2}^{\prime})\varphi-g_{2}\tilde{\varphi}\bigr)\Big]\psi_{k}
\label{AddTerm}%
\end{equation}
to the integrand of $\mathcal{J}_{z}$ (\ref{H-quasi}). Note that this term
arises formally if one replaces each causal ordered $\tilde{n}n$-pair in
$\mathcal{J}_{z}$ by the sum over $\psi_{k}\psi_{k}/2$-pairs. Here comes a new
symmetry into play: the additional term (\ref{AddTerm}) is trivially invariant
under any permutation of the $D$ ghost fields $\psi_{k}$, i.e.,  we have symmetry
under the permutation (or symmetric) group $S_{D}$. However, since in general
$\sum_{k=1}^{D}\psi_{k}\neq0$, this representation is reducible. Hence,
it is more useful to introduce new ghost fields $(\chi_{1},\ldots,\chi_{D+1})$
with constraint $\sum_{\alpha=1}^{D+1}\chi_{\alpha}=0$, and $\sum_{k=1}%
^{D}\psi_{k}^{2}=\sum_{\alpha=1}^{D+1}\chi_{\alpha}^{2}=:\chi^{2}$. This is
easily achieved by using $D+1$ Potts--spin-vectors $\vec{e}^{(\alpha)}%
=(e_{k}^{(\alpha)})$ directed to the corners of a $D$-dimensional simplex.
The spin-vectors have the usual properties: $\sum_{\alpha=1}^{D+1}e_{k}
^{(\alpha)}=0$, $\sum_{\alpha=1}^{D+1}e_{i}^{(\alpha)}e_{k}^{(\alpha)}=\delta_{ik}$,
$\sum_{k=1}^{D}e_{k}^{(\alpha)}e_{k}^{(\beta)}=\delta_{\alpha\beta}-1/(D+1)$.
Hence, the relation between the old and the new ghosts are given by
$\chi_{\alpha}=\sum_{k=1}^{D}e_{k}^{(\alpha)}\psi_{k}$. Now, we have symmetry
under the permutation group $S_{D+1}$ of permutations of the $(D+1)$ ghost
fields $\chi_{\alpha}$, and this representation is irreducible.
\begin{figure}[ptb]
\includegraphics[width=5cm]{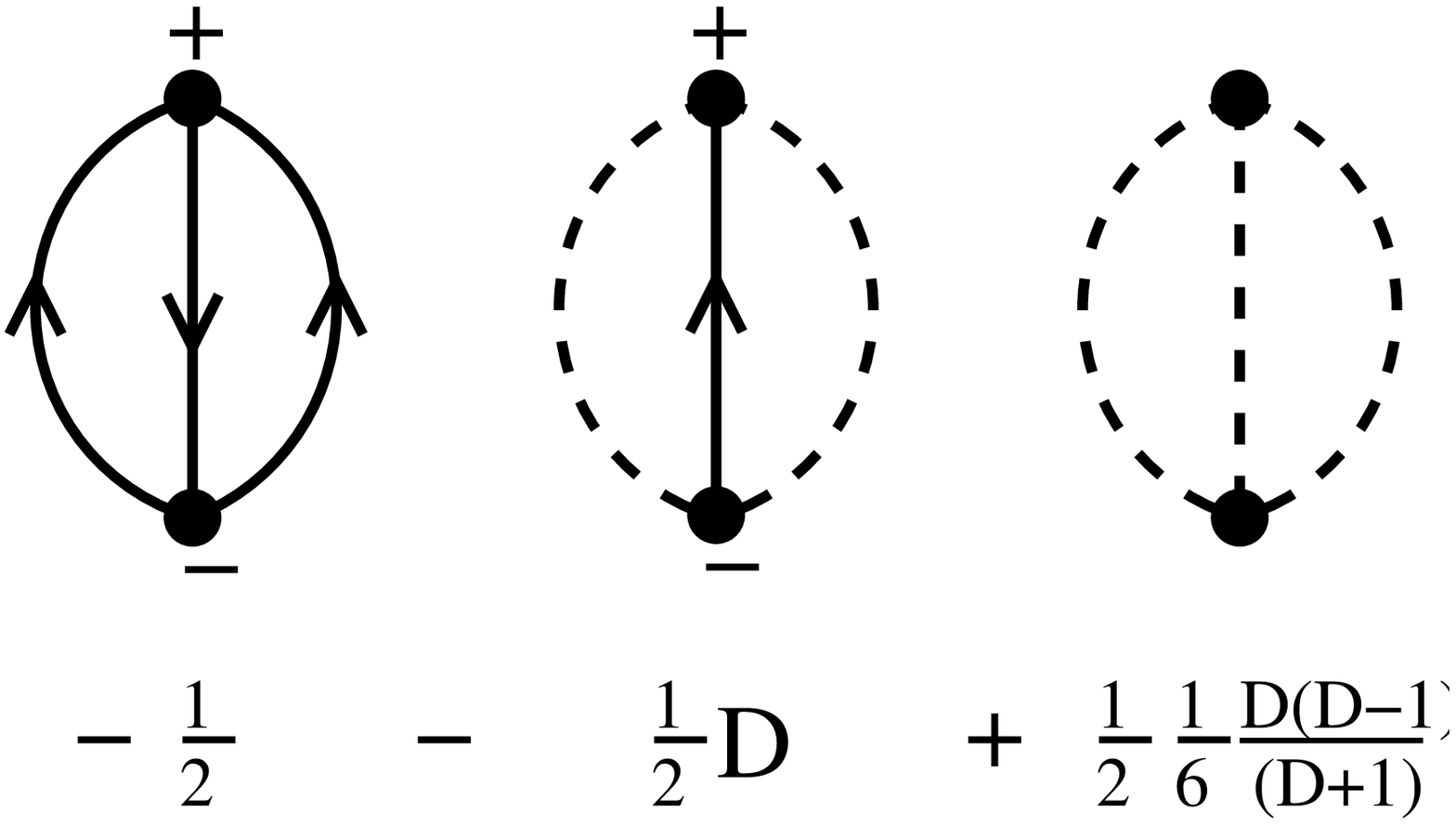}\caption{Basic diagrams for the
cancellation of coupled non-causal loops by $D=-2$ bosonic ghosts.}%
\label{fig:bosonicGhosts}%
\end{figure}

Inspection shows that the ghosts also work in multi-loop diagrams provided
that the non-causal loops are separated from each other in these
diagrams~\cite{JaLy94}. However, as long as $g_{2}^{\prime}$ is not zero (note that $g_{2}$ is always greater than zero because otherwise only diagrams without loops are generated),
non-separated non-causal loops arise, see the first diagram in
Fig.~\ref{fig:bosonicGhosts}. The cancellation
requires a permutation-symmetric irreducible interaction $\chi^{3}%
=\sum_{\alpha=1}^{D+1}\chi_{\alpha}^{3}$ of the $(D+1)$ ghosts, see the third
diagram in Fig.~\ref{fig:bosonicGhosts}. Using these new ghosts, the quasi
static Hamiltonian becomes
\begin{align}
\mathcal{H}  &  =\int d^{d}x\Big\{\tilde{\varphi}\bigl(\tau-\nabla
^{2}\bigr)\varphi+\frac{\rho}{2}\tilde{\varphi}^{2}+h\tilde{\varphi
}\nonumber\\
&  +\frac{1}{2}\bigl(\tau\chi^{2}+(\nabla\chi)^{2}\bigr)+\frac{g_{0}}{6}%
\tilde{\varphi}^{3}+\frac{g_{1}}{2}\tilde{\varphi}\bigl[2\tilde{\varphi
}{\varphi}+\chi^{2}\bigr]\nonumber\\
&  +\frac{1}{6}\bigl[3\tilde{\varphi}(g_{2}^{\prime}\tilde{\varphi}%
-g_{2}\varphi)\varphi+3(g_{2}^{\prime}\tilde{\varphi}-g_{2}\varphi)\chi
^{2}\nonumber\\
&  +\sqrt{g_{2}^{\prime}g_{2}}\chi^{3}\bigr]\Big\}\,. \label{H_1}%
\end{align}
Perturbation theory with this Hamiltonian is no longer burdened with additional rules. It will serve as the vantage point of our RG
calculations. As it stands, it is general enough to capture both the swollen
phase and the collapse transition. As we have shown in mean field theory, the
collapse transition corresponds to vanishing $\tau$, $\rho$, and $h$. Swollen
RBPs correspond to vanishing $\tau$ and $h$, but positive and finite $\rho
$.

The Hamiltonian (\ref{H_1}) is form-invariant under three transformations of the fields.
Therefore, three parameters of the Hamiltonian are
redundant. One of these transformations, the mixing $\varphi\rightarrow
\varphi+\kappa\tilde{\varphi}$, $\tilde{\varphi}\rightarrow\tilde{\varphi}$,
we have already used to eliminate the gradient term $(\nabla\tilde{\varphi
})^{2}$. The second of these transformations, the rescaling $\varphi\rightarrow\lambda\varphi$, $\tilde{\varphi
}\rightarrow\lambda^{-1}\tilde{\varphi}$ can be used either to identify
coupling-constants $g_{2}^{\prime}=g_{2}$, or to transform $g_{2}$ to one and
use only scaling-invariant quantities. Via the third transformation, the shift
$\varphi\rightarrow\varphi+\gamma$, $\tilde{\varphi}\rightarrow\tilde{\varphi
}$, either $\tau$ or $\rho$ can be transformed away. At this point, a word of caution is in order. Using these transformations to eliminate parameters from the field theoretic functional, one is well advised to make sure that non of the parameters $\kappa$,
$\lambda$, or $\gamma$ featured in the transformations is singular. Otherwise, parameters eliminated from the unrenormalized theory will have to re-emerge in the renormalization procedure. This is no problem per se, but it is a fact that can be easily overlooked, and if so, will lead to ill-defined renormalization schemes.

Before moving on to our actual RG calculation, we find it worthwhile to comment on the renormalizability of $\mathcal{H}$. Simple power counting shows that the ghosts have the same dimensionality as the fields $\tilde{\varphi}$ and ${\varphi}$, namely $\chi_{\alpha}\sim\mu^{(d-2)/2}$. For the swollen phase,  the coupling constants $g_{0}$, $g_{1}$, and $g_{2}^{\prime}$ are irrelevant and hence can and should be set equal to zero. Then one can easily ascertain that the remaining $\mathcal{H}$ contains all the relevant terms generated under renormalization, and hence $\mathcal{H}$ is renormalizable as far as the swollen phase is concerned. For the collapse transition, the situation is more intricate. Simple inspection by means of power counting lends credence to the renormalizability of $\mathcal{H}$. However, one has to be more careful here, because the way the various $g$'s appear in multiple places, i.e.,
a given $g$ may appear as a factor of different monomials
of the fields, viz. in couplings amongst the ghost, in couplings amongst the
primary fields $\varphi$ and $\tilde{\varphi}$, and in couplings of the
primary fields and the ghosts. Does this spoil renormalizability? The answer is clearly {\em no} because we know for certain that $\mathcal{H}$ is renormalizable by virtue of its equivalence in the quasi-static limit to the renormalizable
dynamic functional $\mathcal{J}_{z}$ which is renormalizable. Hence, there must exist some hidden
symmetry that masks the renormalizability of
$\mathcal{H}$. Once revealed, this underlying symmetry will provide for relations between
different vertex-functions. We will show shortly that this is the symmetric group
$S_{D+2}$ (not only the permutation symmetry $S_{D+1}$ of the $D+1$ ghosts alone) of  the permutation of $(D+2)$ field combinations. First, however, we will look briefly at a $1$-loop calculation that underpins and exemplifies the considerations just presented.

\subsection{1-loop diagrams with ghosts}

The elements of our diagrammatic perturbation expansion, the propagators, the correlators, and
the vertices, are listed in Fig.~(\ref{fig:propas}) and Fig.~(\ref{fig:vertexe}), respectively. For the time being, we focus here just on the decorations of Feynman diagrams, i.e.,  the combinations of coupling-constants and symmetry-factors of the diagrams without the integrations over loop-momenta.  We list the relevant $1$-loop diagrams, writing them in a form that makes evident the cancellations in the limit $D\rightarrow-2$. For the
tadpole-diagrams, Fig.~(\ref{fig:tadpoles}), we find
\begin{subequations}
\begin{align}
1a) &  :\qquad g_{2}+\frac{D}{2}g_{2}\rightarrow0\,,\\
1b) &  :\qquad-\bigl[g_{1}+(g_{1}+g_{2}^{\prime})\bigr]-\frac{D}{2}%
(g_{1}+g_{2}^{\prime})\rightarrow-g_{1}\,.
\end{align}
The selfenergy-diagrams, Fig.~(\ref{fig:selfenergy}), yield%
\end{subequations}
\begin{subequations}
\begin{align}
2a):\qquad &  g_{2}^{2}+\frac{D}{2}g_{2}^{2}\rightarrow0\,,\\
2b):\qquad &  -\frac{1}{2}g_{2}(2g_{1}+g_{2}^{\prime})-g_{2}\bigl[g_{1}%
+(g_{1}+g_{2}^{\prime})\bigr]\nonumber\\
&  -\frac{D}{2}g_{2}(g_{1}+g_{2}^{\prime})\rightarrow-2g_{1}g_{2}-\frac{1}%
{2}g_{2}g_{2}^{\prime}\,,\\
2c):\qquad &  -g_{0}g_{2}+\bigl[g_{1}+(g_{1}+g_{2}^{\prime})\bigr]^{2}%
\nonumber\\
&  +\frac{D}{2}(g_{1}+g_{2}^{\prime})^{2}\rightarrow-g_{0}g_{2}+3g_{1}%
^{2}+2g_{1}g_{2}^{\prime}\,.
\end{align}
In the same way we obtain the decorations of the vertex-diagrams,
Figs.~(\ref{fig:vertexo}) to (\ref{fig:vertexc}),
\end{subequations}
\begin{subequations}
\begin{align}
3a):\qquad &  2g_{2}^{3}+Dg_{2}^{3}\rightarrow0\,,\\
3b):\qquad &  -2g_{2}^{2}(2g_{1}+g_{2}^{\prime})-2g_{2}^{2}\bigl[g_{1}%
+(g_{1}+g_{2}^{\prime})\bigr]\nonumber\\
&  -Dg_{2}^{2}(g_{1}+g_{2}^{\prime})\rightarrow-6g_{1}g_{2}^{2}-2g_{2}%
^{2}g_{2}^{\prime}\,,\\
3c):\qquad &  2g_{2}(2g_{1}+g_{2}^{\prime})^{2}-2g_{0}g_{2}^{2}+2g_{2}%
\bigl[g_{1}+(g_{1}+g_{2}^{\prime})\bigr]^{2}\nonumber\\
&  +Dg_{2}(g_{1}+g_{2}^{\prime})^{2}\rightarrow14g_{1}^{2}g_{2}+12g_{1}%
g_{2}g_{2}^{\prime}\nonumber\\
&  \qquad\qquad\qquad\qquad+2g_{2}g_{2}^{\prime2}-2g_{0}g_{2}^{2}\,,\\
3d):\qquad &  6g_{0}g_{2}(2g_{1}+g_{2}^{\prime})-2\bigl[g_{1}+(g_{1}%
+g_{2}^{\prime})\bigr]^{3}\nonumber\\
&  +2D(g_{1}+g_{2}^{\prime})^{3}\rightarrow12g_{0}g_{1}g_{2}+6g_{0}g_{2}%
g_{2}^{\prime}\nonumber\\
&  \qquad\qquad-14g_{1}^{3}-18g_{1}^{2}g_{2}g_{2}^{\prime}-6g_{1}g_{2}%
^{\prime2}\,.
\end{align}
\begin{figure}[ptb]
\includegraphics[width=5cm]{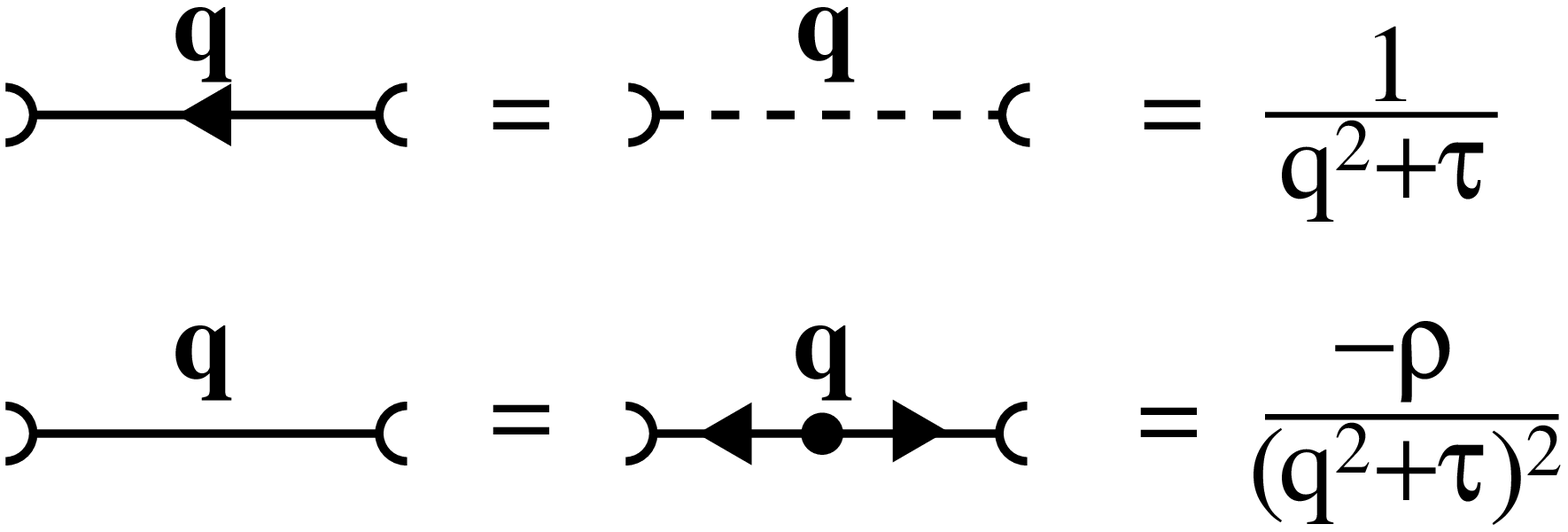}\caption{Propagators and
correlators.}%
\label{fig:propas}%
\end{figure}
\begin{figure}[ptb]
\includegraphics[width=7cm]{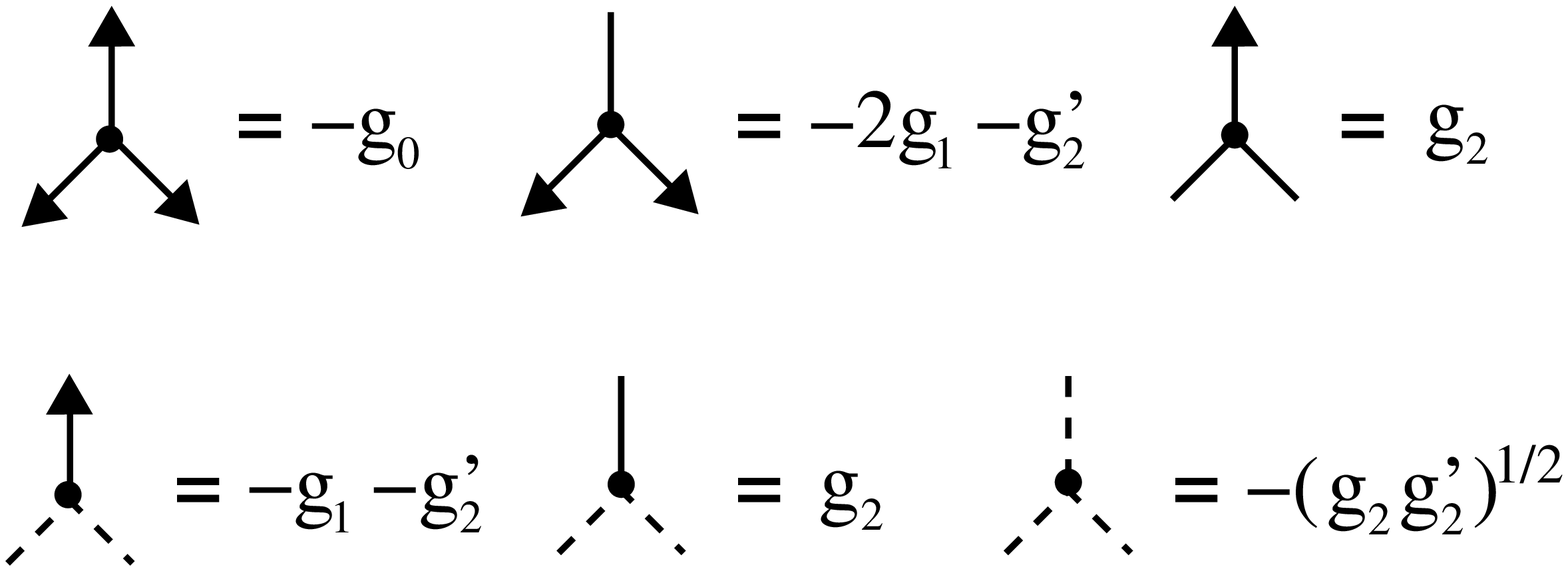}\caption{Vertices.}%
\label{fig:vertexe}%
\end{figure}
\begin{figure}[ptb]
\includegraphics[width=5cm]{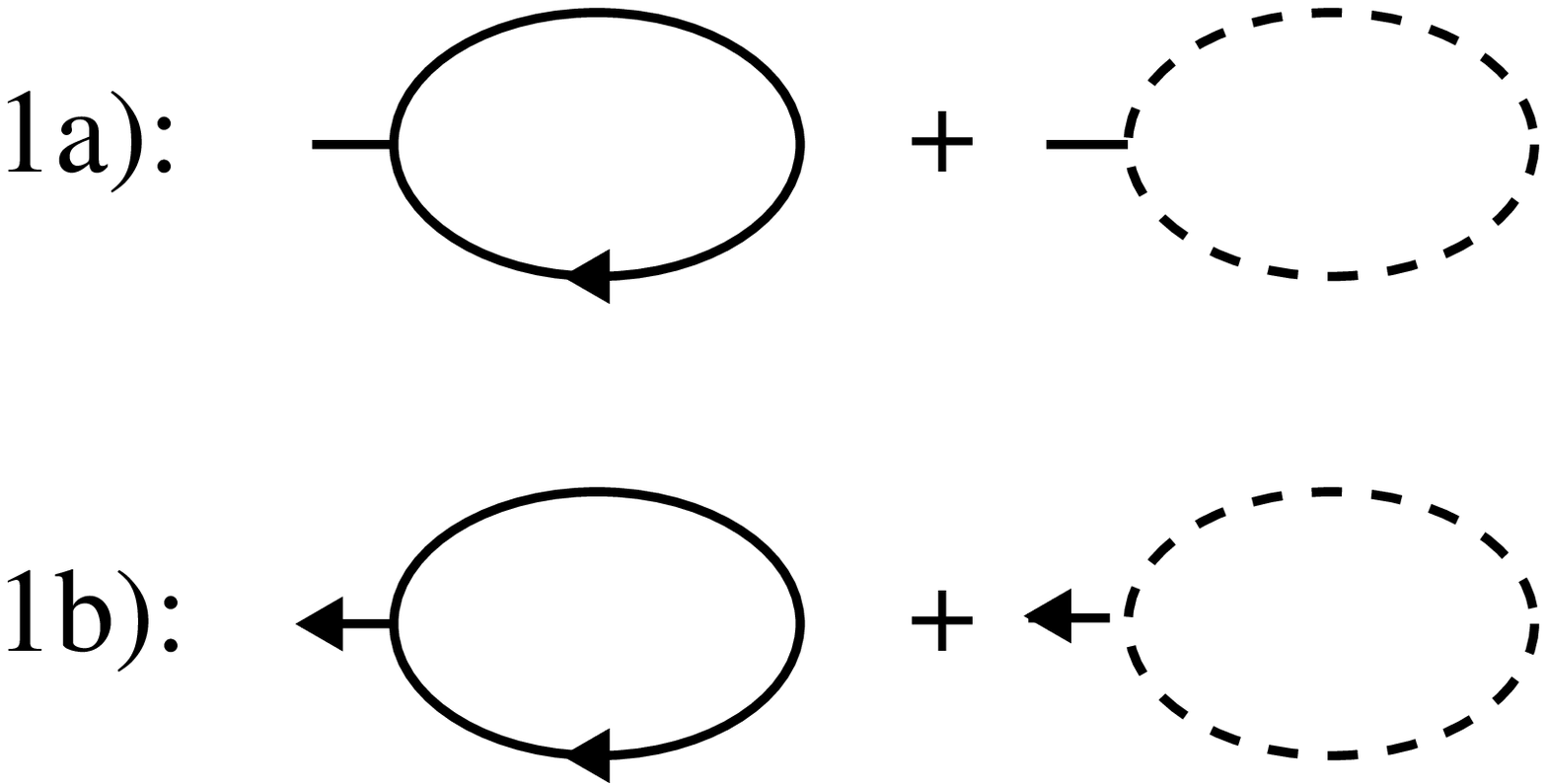}\caption{1-loop tadpole diagrams.}%
\label{fig:tadpoles}%
\end{figure}
\begin{figure}[ptb]
\includegraphics[width=7cm]{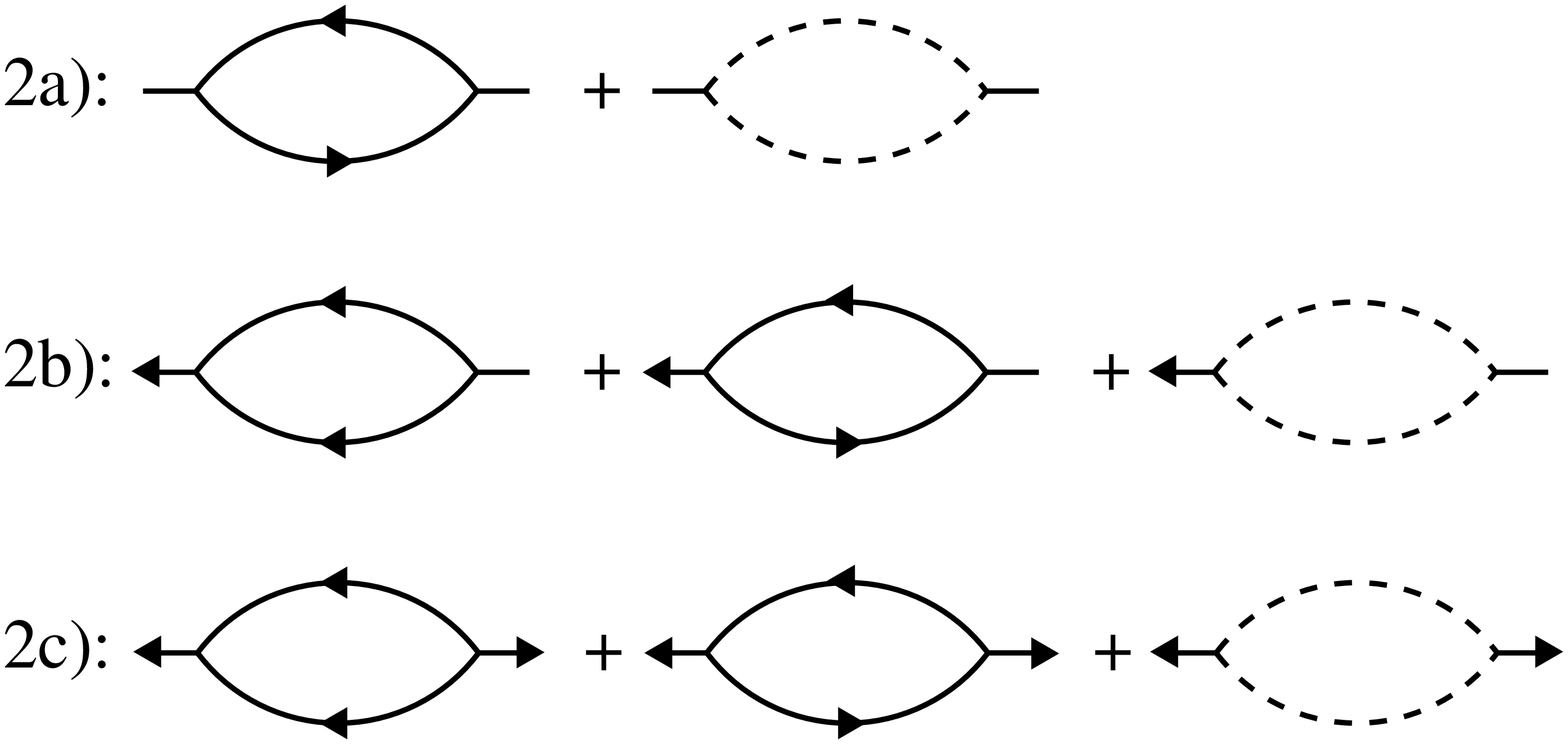}\caption{1-loop self-energy diagrams.}%
\label{fig:selfenergy}%
\end{figure}
\begin{figure}[ptb]
\includegraphics[width=4cm]{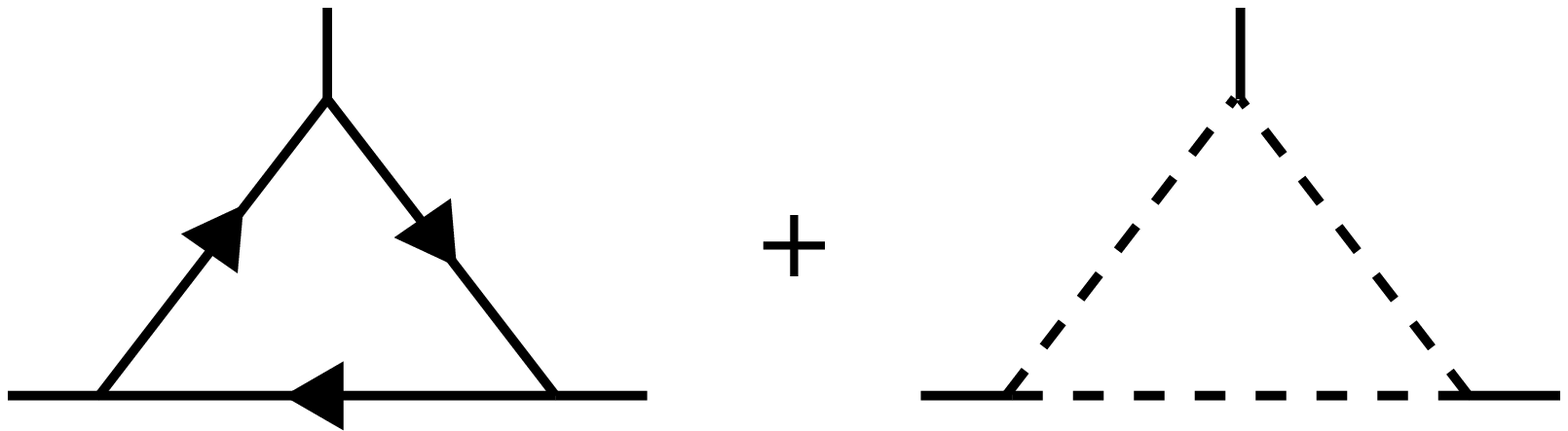}\caption{1-loop vertex diagrams a.}%
\label{fig:vertexo}%
\end{figure}
\begin{figure}[ptb]
\includegraphics[width=6cm]{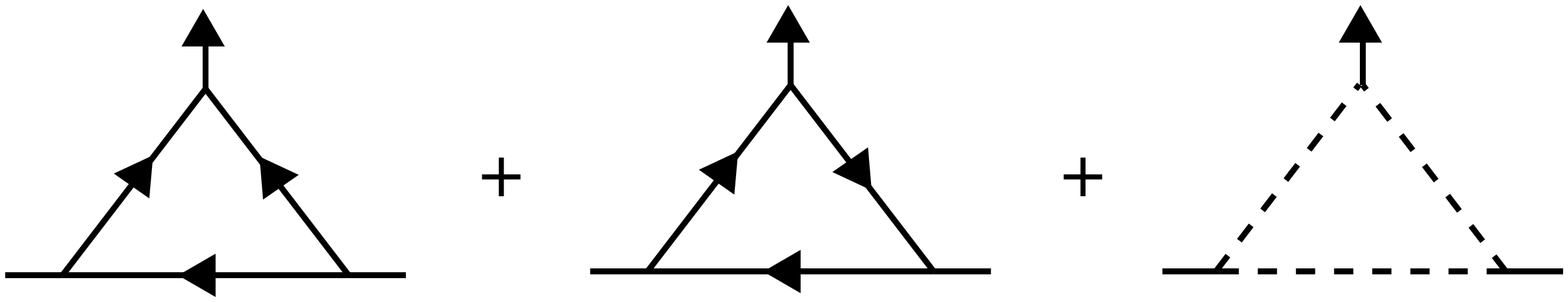}\caption{1-loop vertex diagrams b.}%
\label{fig:vertexa}%
\end{figure}
\begin{figure}[ptb]
\includegraphics[width=4cm]{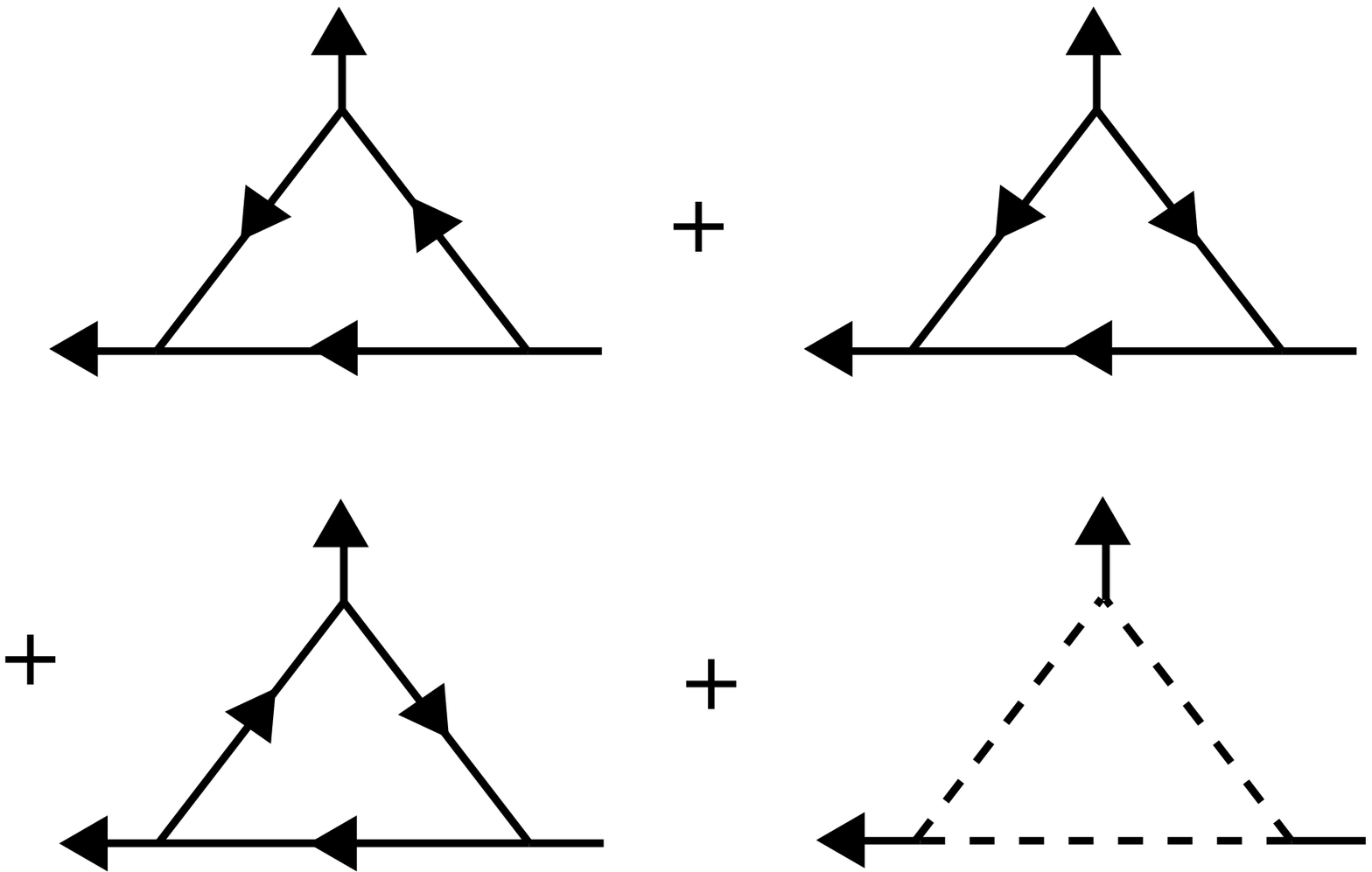}\caption{1-loop vertex diagrams c.}%
\label{fig:vertexb}%
\end{figure}
\begin{figure}[ptb]
\includegraphics[width=6cm]{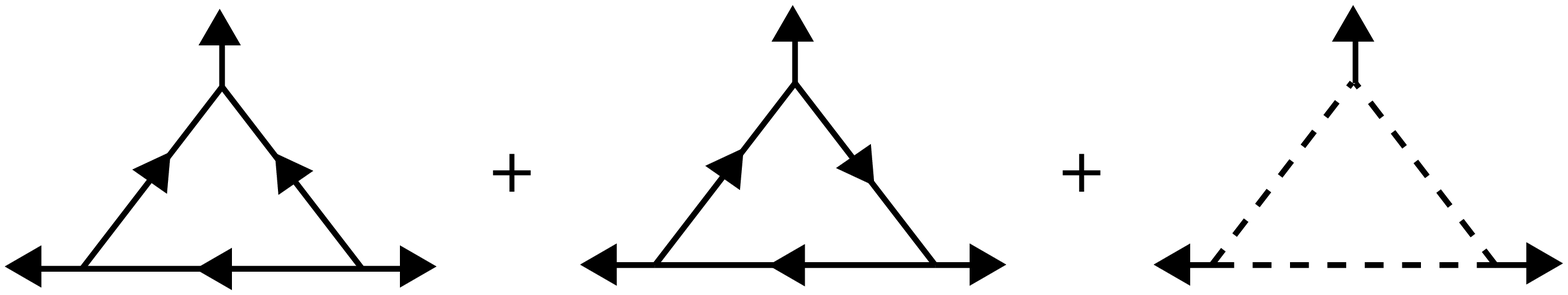}\caption{1-loop vertex diagrams d.}%
\label{fig:vertexc}%
\end{figure}
\begin{figure}[ptb]
\includegraphics[width=5cm]{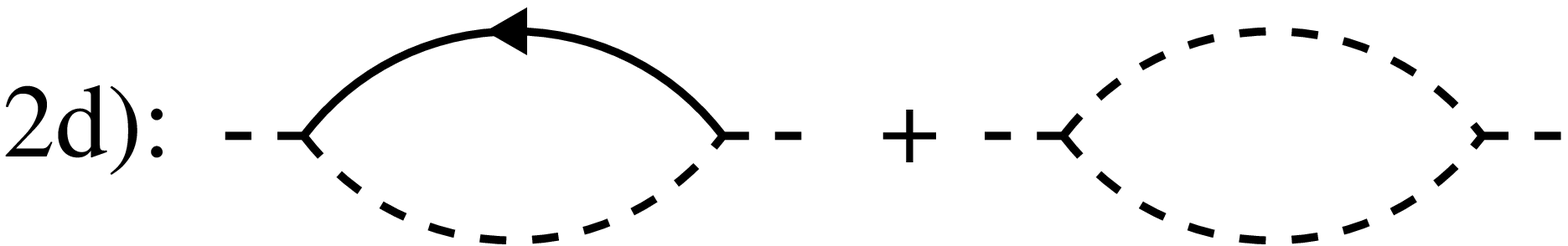}\caption{1-loop self-energy diagrams of
ghosts.}%
\label{fig:ghostselfenergy}%
\end{figure}

Of course, the cancellation of non-causal loops, see
Fig.~(\ref{fig:cancellation1Loop}), should occur also in higher loop-orders.
Hence, not only the propagator and the ghost-correlator must be equal but also
the full Greens functions $\langle\varphi(\mathbf{r})\tilde{\varphi
}(\mathbf{0})\rangle$ and $\langle\chi(\mathbf{r})\chi(\mathbf{0})\rangle$.
Therefore, the 1-loop self-energy diagrams  $2b)$ shown in Fig.~(\ref{fig:selfenergy}) must be equal to the 1-loop self-energy diagrams $2d)$ of ghost  shown in Fig.~(\ref{fig:ghostselfenergy}):
\end{subequations}
\begin{align}
2d)  &  =-2g_{2}(g_{1}+g_{2}^{\prime})+\frac{1}{2}\Big(\frac{D-1}%
{D+1}\Big)g_{2}g_{2}^{\prime}\nonumber\\
&  \rightarrow-2g_{1}g_{2}-\frac{1}{2}g_{2}g_{2}^{\prime}=2b)\,.
\end{align}
Hence, the $\sim\chi^{3}$  self-interaction of ghosts featured in $\mathcal{H}$ is needed already at
$1$-loop order to guarantee the equality of self-energies.

\subsection{Hidden symmetry and relation to other models}

Now, we come back to the search for the symmetry that ensures the
renormalizability of the Hamiltonian $\mathcal{H}$, Eq.~(\ref{H_1}). At first
glance, this Hamiltonian only has the permutation-symmetry $S_{D+1}$ of the
$(D+1)$ ghost-fields $(\chi_{\alpha})$. Next we use the form-invariance of the
Hamiltonian under a rescaling of the original fields%
\begin{equation}
\varphi\rightarrow\lambda\varphi\,,\qquad\tilde{\varphi}\rightarrow
\lambda^{-1}\tilde{\varphi}\,,\label{Resc1}%
\end{equation}
which is compensated for and hence becomes a scaling symmetry when augmented by the following redefinition of
parameters:
\begin{align}
g_{0} &  \rightarrow\lambda^{3}g_{0}\,,\quad g_{1}\rightarrow\lambda
g_{1}\,,\quad g_{2}^{\prime}\rightarrow\lambda g_{2}^{\prime}\,,\nonumber\\
g_{2} &  \rightarrow\lambda^{-1}g_{2}\,,\quad\rho\rightarrow\lambda^{2}%
\rho\,,\quad h\rightarrow\lambda h\,.\label{Resc2}%
\end{align}
Under the choice $\lambda=\sqrt{g_{2}^{\prime}/g_{2}}$, which is possible as
long as $g_{2}^{\prime}\neq0$, we gain the equality $g_{2}^{\prime}=g_{2}$.
Now, it is easy to show that the Hamiltonian $\mathcal{H}$ is invariant for
each $\alpha=1,\ldots,D+1$ under the mirror-transformations
\begin{align}
\tilde{\varphi} &  \rightarrow\tilde{\varphi}\,,\qquad\qquad\quad
\varphi\rightarrow(\varphi-\tilde{\varphi})-\chi_{\alpha}\,,\nonumber\\
\chi_{\alpha} &  \rightarrow-\chi_{\alpha}-2\tilde{\varphi}\,,\quad\chi
_{\beta}\rightarrow(\chi_{\beta}-\chi_{\alpha})-\tilde{\varphi}%
\,,\label{mirrtraf}%
\end{align}
for all $\beta\neq\alpha,$ and always in the limit $D\rightarrow-2$. This
invariance ensures, e.g., the equality of the ghost-correlation functions with
the propagator%
\begin{equation}
\langle\chi_{\alpha}\chi_{\beta}\rangle=\Big(\delta_{\alpha\beta}-\frac
{1}{D+1}\Big)\langle\varphi\tilde{\varphi}\rangle
\end{equation}
that we have demonstrated explicitly to $1$-loop order above. The mirror-transformations which
mix original fields with ghosts complete the permutation-symmetries of the
ghosts to the full symmetry-group $S_{D+2}$.

To make this hidden symmetry more transparent,
we define a new order parameter field with $(D+3)$ components: $s_{0}%
=\tilde{\varphi}$, $s_{1}=-{\varphi}$, and for $\mu\geq2$: $s_{\mu}=\chi
_{\mu-1}-(\tilde{\varphi}-{\varphi})/(D+1)$. With $s^{k}:=\sum_{\mu=0}%
^{n}s_{\mu}^{k}$, where $n=D+2$, we have $s^{1}=0$, and in the limit
$n\rightarrow0$
\begin{subequations}
\begin{align}
s^{2} &  =2\tilde{\varphi}{\varphi+}\chi^{2}\,,\\
s^{3} &  =3\tilde{\varphi}(\tilde{\varphi}-{\varphi}){\varphi+3(\tilde
{\varphi}-{\varphi})+}\chi^{3}\,.
\end{align}
Using this order parameter, it is easy to see that the Hamiltonian can be
written as%
\end{subequations}
\begin{align}
\mathcal{H}_{\mathrm{aP}} &  =\int d^{d}x\Big\{\frac{1}{2}\bigl(\tau
s^{2}+(\nabla s)^{2}\bigr)+\frac{\rho}{2}s_{0}^{2}+hs_{0}\nonumber\\
&  \qquad\qquad+\frac{g_{0}}{6}s_{0}^{3}+\frac{g_{1}}{2}s_{0}s^{2}+\frac
{g_{2}}{6}s^{3}\Big\}\,,\label{H_2}%
\end{align}
which is identical to the Hamiltonian of Eq.\ (\ref{H_1}) in the limit
$n\rightarrow0$. It is therefore equivalent to our original dynamical model.
The Hamiltonian $\mathcal{H}_{\mathrm{aP}}$ describes the field theory of the
asymmetric $(n+1)$-state Potts model. The previously hidden symmetry is now the symmetry
$S_{n}$ of permutations of the $n$ fields $(s_{1},\ldots s_{n})$. As mentioned
in the introduction, the established theories for
RBPs~\cite{LuIs78,HaLu81,Con83} are mainly based on the asymmetric Potts
model, and the Hamiltonian~(\ref{H_2}) therefore establishes the connection
with these theories. Note, furthermore, that for $g_{0}=g_{1}=0$ the
Hamiltonian~(\ref{H_2}) describes the symmetric $(n+1)$-state Potts model with
a linear and quadratic (so-called hard) symmetry breaking. The interaction
represented by the third-order terms has $S_{n+1}$-symmetry and yields the
field theory of percolation in the limit $n\rightarrow0$. There exists another
connection: one can show that for $4g_{0}=2g_{1}=-g_{2}$ the Hamiltonian
$\mathcal{H}_{\mathrm{aP}}$ decomposes in a sum of $n$ uncoupled Hamiltonians
each describing the Yang-Lee edge problem. The choice of these special
combinations of coupling-constants yield important checks for higher order
calculations \cite{JaSt-prep}.

Now, we turn to the case that $g_{2}^{\prime}$ is zero where we cannot rescale
the fields to attain $g_{2}^{\prime}=g_{2}$. As we will show, $g_{2}^{\prime}$
becomes zero at the fixed point of our model for the collapse, and it is irrelevant
for the model in the swollen phase of the RBP. Hence, the case $g_{2}^{\prime
}=0$ is important in general for the statistics of branched polymers. Now the
third order coupling $\sim\chi^{3}$ of the ghosts in $\mathcal{H}$,
Eq.~(\ref{H_1}), vanishes. The ghosts appear only quadratic, and we can
 integrate them out formally producing a ghost-determinant raised to the power $(-D/2)$.
Taking the limit $D\rightarrow-2$, this determinant can be reimported into the
Hamiltonian by introducing a pair $(\bar{\psi},\psi)$ of anticommuting
fermionic ghost-fields. The Hamiltonian becomes
\begin{align}
\mathcal{H}_{\mathrm{ss}}  &  =\int d^{d}x\Big\{\tilde{\varphi}\bigl(\tau
-\nabla^{2}\bigr)\varphi+\frac{\rho}{2}\tilde{\varphi}^{2}+h\tilde{\varphi
}\nonumber\\
&  +\bar{\psi}\bigl(\tau-\nabla^{2}+g_{1}\tilde{\varphi}-g_{2}\varphi
\bigr)\psi\nonumber\\
&  +\frac{g_{0}}{6}\tilde{\varphi}^{3}+g_{1}\tilde{\varphi}^{2}{\varphi}%
-\frac{g_{2}}{2}\tilde{\varphi}\varphi^{2}\Big\}\,. \label{H-super}%
\end{align}
Introducing Grassmannian anticommuting super-coordinates $\theta$,
$\bar{\theta}$ with integration rules {$\int d\theta\,1={\int d\bar{\theta
}\,1=}0$, $\int d\theta\,\theta={\int d\bar{\theta}\,\bar{\theta}=}1,$ and
defining a super-field $\Phi(\mathbf{r},\bar{\theta},\theta)=i\varphi
(\mathbf{r})+\bar{\theta}\psi(\mathbf{r})+\bar{\psi}(\mathbf{r})\theta
+i\bar{\theta}\theta\tilde{\varphi}(\mathbf{r})$, the Hamiltonian
}$\mathcal{H}_{\mathrm{ss}}$ takes the form%
\begin{align}
\mathcal{H}_{\mathrm{ss}}  &  =\int d^{d}xd\bar{\theta}d\theta\,\Big\{\frac
{1}{2}\Phi\Big(\tau-\nabla^{2}-\rho\partial_{\bar{\theta}}\partial_{\theta
}\Big)\Phi+ih\Phi\nonumber\\
&  +i\Big(\frac{g_{2}}{6}\Phi^{3}+\frac{g_{1}}{2}\Phi^{2}(\partial
_{\bar{\theta}}\partial_{\theta}\Phi)-\frac{g_{0}}{6}\Phi(\partial
_{\bar{\theta}}\partial_{\theta}\Phi)^{2}\Big)\Big\}\,. \label{H-super1}%
\end{align}
This Hamiltonian shows Becchi--Rouet--Stora\thinspace(BRS)-symmetry
\cite{BRS75,ZJ02}, i.e., $\mathcal{H}_{\mathrm{ss}}$ is invariant under a
super-translation {$\theta\rightarrow\theta+\varepsilon,$ $\bar{\theta
}\rightarrow\bar{\theta}+\bar{\varepsilon}$. Moreover, if the control
parameter }$\rho$ is positive and finite, i.e., if we consider the problem of
swollen RBPs, $\rho$ can be reset by a scale transformation of the
super-coordinates to $2$. The super-coordinates get a dimension $\sim
\mu^{-1}$ equal to the dimension of the spatial coordinates, and the
derivatives combine to a super-Laplace operator $\nabla^{2}+\rho\partial_{\bar{\theta}%
}\partial_{\theta}\rightarrow\nabla^{2}+2\partial_{\bar{\theta}}%
\partial_{\theta}=:\square$. As we have shown above, the coupling constants
$g_{0}$ and $g_{1}$ become irrelevant and hence can be neglected in which case
the Hamiltonian takes the super-Yang-Lee form
\begin{equation}
\mathcal{H}_{\mathrm{sYL}}=\int d^{d}xd\bar{\theta}d\theta\,\Big\{\frac{1}%
{2}\Phi\Big(\tau-\square\Big)\Phi+i\frac{g}{6}\Phi^{3}+ih\Phi\Big\}\,,
\end{equation}
where we have set $g_{2}=g$. The Hamiltonian $\mathcal{H}_{\mathrm{sYL}}$
has, besides the super-translation invariance, super-rotation invariance.
Now dimensional reduction can be used to reduce the problem to the normal
Yang-Lee problem in two lesser dimensions. This establishes the connection
between our model and the work of Parisi and Sourlas~\cite{PaSo81} on swollen RBPs.

Before moving on to the core of our RG analysis, we would like to highlight
the following implication of our symmetry considerations for the collapse
transition. We will see later on that $g_{2}^{\prime}$ vanishes at the RG
fixed point describing the $\theta$-transition. Thus, this transition is
associated with BRS symmetry which is in contrast to the swollen phase which
is associated with full super-symmetry. The BRS symmetry indicates that the statistics of the RBPs is dominated by tree configurations.This fact can be understood, for example, by using Cardy's presentation \cite{Ca2003} of the work of Brydges and Imbrie \cite{BrIm03}. Cardy reformulates their model of swollen RBPs in $d$ dimensions (which is exactly reducible to the problem of the universal repulsive gas singularity in $d-2$ dimensions which, in turn, belongs to the same universality class as the Yang-Lee problem) in a fully supersymmetric way. If one adds an attracting potential between the monomers of the tree-polymers that can lead eventually to the collapse of the RBPs, the rotational supersymmetry is lost, and with it dimensional reduction. However, BRS symmetry is retained, and this symmetry is indeed the vehicle that reduces all configurations to trees.
Another route to understand the connection between BRS symmetry and trees lies in a dynamical calculation. At first, this may sound somewhat surprising because BRS symmetry is a feature of the quasi-static Hamiltonian at the collapse fixed point. However, a calculation \cite{JaSt-prep} of the fractal dimension of the minimal path from the original dynamic model (\ref{Jz}) with $g_2^{\prime}=0$ clearly shows that the backbone of the RBPs is topologically $1$-dimensional. Thus, asymptotically large RBPs at the $\theta$-transition have the topology of trees.

\section{Renormalization and the renormalization group}

\label{sec:renColTran}

Now, we turn to the core of our RG analysis. As announced above, we will base
our discussion on the Hamiltonian $\mathcal{H}$ of Eq.~(\ref{H_1}). Likewise,
we could use $\mathcal{H}_{\mathrm{aP}}$ with the limit $n\rightarrow0$ which,
as we have shown above, is equivalent to $\mathcal{H}$. For our discussion
here, we choose $\mathcal{H}$ over $\mathcal{H}_{\mathrm{aP}}$ because we feel
that the relation of the former to the original GEP is somewhat more intuitive
than that of the latter. Actual diagrammatic calculations in higher
loop-orders, however, are better to handle when $\mathcal{H}_{\mathrm{aP}}$
instead of $\mathcal{H}$ is used. The renormalization-group functions\ that
feed into our RG analysis for RBPs stem from a renormalized field theoretic
calculation for the asymmetric Potts model that we performed recently. Details
of this work will be presented elsewhere~\cite{JaSt-prep}.

\subsection{The renormalization scheme}

Our main focus here lies on the collapse transition, i.e., we are mainly
interested in the case that the control parameters $\tau$ and $\rho$ take
critical values (zero in mean-field theory) where the correlation length
diverges, and correlations between different polymers vanish. Via the equation
of state this implies the critical value of $h$. The principal objects of the
perturbation theory are the superficially UV-divergent vertex functions
$\Gamma_{\tilde{k},k}$ which consist of irreducible diagrams with $\tilde{k}$
and $k$ amputated legs of $\tilde{\varphi}$ and $\varphi$, respectively, as
functions of the wave vector $\mathbf{q}$. The UV-divergences are then handled
via a renormalization scheme that introduces counter terms which absorb said
divergences. For our calculations, we use minimal renormalization, i.e.,
dimensional regularization and minimal subtraction in conjunction with the
$\varepsilon$-expansion about $d=6$ dimensions ($\varepsilon=6-d$). Our
renormalization scheme leading from bare to renormalized quantities reads
\begin{subequations}
\label{Reno}%
\begin{align}
(\tilde{\varphi},\varphi,\chi) &  \rightarrow(\mathring{\tilde{\varphi}%
},\mathring{\varphi},\mathring{\chi})=Z^{1/2}(\tilde{\varphi},\varphi
+K\tilde{\varphi},\chi)\,,\label{Reno1}\\
\underline{\tau} &  \rightarrow\underline{\mathring{\tau}}=Z^{-1}%
\underline{\underline{Z}}\cdot\underline{\tau}+\underline{\mathring{\tau}}%
_{c}\,,\label{Reno2}\\
h &  \rightarrow\mathring{h}=Z^{-1/2}(h+\frac{1}{2}G_{\varepsilon}^{1/2}%
\mu^{-\varepsilon/2}\underline{\tau}\cdot\underline{\underline{A}}%
\cdot\underline{\tau})\nonumber\\
&  \qquad\qquad\qquad+\underline{\mathring{C}}_{c}\cdot{\underline{\tau}%
+}\mathring{h}_{c}{\,,}\label{Reno3}\\
G_{\varepsilon}^{1/2}g{_{\alpha}} &  \rightarrow G_{\varepsilon}%
^{1/2}\mathring{g_{\alpha}}=Z^{-3/2}(u_{\alpha}+B_{\alpha})\mu^{\varepsilon
/2}\,,\label{Reno4}%
\end{align}
where $G_{\varepsilon}$ is
a convenient numerical factor which we chose here to be $G_{\varepsilon}%
=\Gamma(1+\varepsilon/2)/(4\pi)^{d/2}$. Note, however, that all choices with
$(4\pi)^{3}G_{\varepsilon}=1+O(\varepsilon)$ work equally well since their
differences only amount to a finite rescaling of the momentum scale $\mu$. We
introduce the two-dimensional control-vector $\underline{\tau}=(\rho,\tau)$,
and $(g{_{\alpha}})=(g_{0},g_{1},g_{2}^{\prime},g_{2})$. In a theory
regularized by means of a large momentum-cutoff $\Lambda$, the additive
non-universal counter terms $\underline{\mathring{\tau}}_{c}$, $\mathring
{h}_{c}$, and $\underline{\mathring{C}}_{c}$ would diverge $\sim\Lambda^{2}$,
$\Lambda^{4-\varepsilon/2}$, and $\Lambda^{2-\varepsilon/2}$. In our
perturbative approach based on dimensional regularisation and minimal
subtraction with $\varepsilon$-expansion, they formally vanish. In minimal
renormalization, all the other counter-terms are expanded into pure
Laurent-series, e.g.,
\end{subequations}
\begin{subequations}
\begin{align}
Z-1 &  =\frac{Z^{(1)}}{\varepsilon}+O(\varepsilon^{-2})\,,\\
K &  =\frac{K^{(1)}}{\varepsilon}+O(\varepsilon^{-2})\,,
\end{align}
and so on, where the residua $Z^{(1)}$, $K^{(1)}$, $\cdots$ of the $\varepsilon$-poles
are pure functions of the dimensionless renormalized coupling-constants
$(u{_{\alpha}})=(u_{0},u_{1},u_{2}^{\prime},u_{2})$. We present the
calculation of all the counter-terms to $1$-loop order in Appendix~\ref{app:1-loopCalc}.

Note that the renormalization scheme~(\ref{Reno}) introduces a counter-term
proportional to $K$ that has no counterpart in the Hamiltonian~(\ref{H_1}).
This counter-term can be viewed as a remnant of the gradient-term proportional
to the redundant parameter $c$ in the original response functional~(\ref{StochFu}) which
we removed from our model via the mixing-transformation stated in Eq.~(\ref{RedTraf}). As
a counter term this term is indispensable, however, because the quadratically
superficial divergent vertex function
\end{subequations}
\begin{equation}
\Gamma_{2,0}(\mathbf{q)=}\Gamma_{2,0}(\mathbf{0)+q}^{2}\Gamma_{2,0}%
^{\prime\prime}(\mathbf{0})+\ldots
\end{equation}
contains an UV-divergent $\Gamma_{2,0}^{\prime\prime}(\mathbf{0})$. This fact
was overlooked by LI \cite{LuIs78} in their calculation, and their
long-standing $1$-loop results are incorrect although, fortunately, the
numeric deviations from the correct $1$-loop results are rather small. It must
be stressed, however, that the omission of this counter term is not just a
technical glitch that affects some numbers. Without this term, renormalization
does not cure the theory from non-primitive divergences and is thus not really
meaningful. In a 1-loop calculation one does not see these non-primitive
divergences explicitly, and hence they are easily overlooked. At higher loop
order, however, they inevitably pop up, and one can see explicitly and in
detail how the theory fails if not renormalized properly.

The alert reader might ask why the different fields are renormalized with the
same renormalization factor $Z$. The fields belong to two different
irreducible representations of the symmetry group $S_{n}$, mathematically
denoted by $\{n\}$ and $\{n-1,1\}$, the trivial and the fundamental
representation, respectively. They should therefore require two independent factors
$Z_{0}$ and $Z_{1}$. In general, this argumentation is correct, and $Z_{0}\neq Z_{1}$ as
long as $n\neq0$ as well as $g_{0}$ or $g_{1}$ are non-zero. In the limit $n\rightarrow0$, however, these renormalization-factors become equal. To demonstrate this, we reduce the order parameter $s=(s_{0}=\tilde{\varphi},s_{1}=-\varphi
,s_{\alpha+1}=\chi_{\alpha}+(\varphi-\tilde{\varphi})/(n-1))$ of the
Hamiltonian $\mathcal{H}_{\mathrm{aP}}$ (\ref{H_2}) into its irreducible
components:
\begin{subequations}
\begin{align}
\phi_{0} &  =\sqrt{\frac{n+1}{n}}s_{0}\in\{n\}\,,\\
\phi_{\nu} &  =s_{\nu}+\frac{1}{n}s_{0}\in\{n-1,1\}\,,
\end{align}
with $\sum_{\nu=1}^{n}\phi_{\nu}=0$, and $s^{2}=\phi_{0}^{2}+\phi^{2}$. The
renormalizations%
\end{subequations}
\begin{subequations}
\begin{align}
\phi_{0} &  \rightarrow\mathring{\phi}_{0}=Z_{0}^{1/2}\phi_{0}\,,\\
\phi_{\nu} &  \rightarrow\mathring{\phi}_{\nu}=Z_{1}^{1/2}\phi_{\nu}\,
\end{align}
lead to
\end{subequations}
\begin{subequations}
\begin{align}
s_{0} &  \rightarrow\mathring{s}_{0}=Z_{0}^{1/2}s_{0}\,,\\
s_{\nu} &  \rightarrow\mathring{s}_{\nu}=Z_{1}^{1/2}s_{\nu}+\frac{1}%
{n}\bigl(Z_{1}^{1/2}-Z_{0}^{1/2}\bigr)s_{0}\,.
\end{align}
We know that these last two renormalizations stay finite in the limit
$n\rightarrow0$ since our primary Hamiltonian (\ref{H_1}) is renormalizable.
Hence,
\end{subequations}
\begin{subequations}
\begin{align}
&  \lim_{n\rightarrow0}Z_{0}=\lim_{n\rightarrow0}Z_{1}=Z\,,\\
&  \lim_{n\rightarrow0}\Big(\frac{(Z_{0}/Z_{1})^{1/2}-1}{n}\Big)=K \, ,
\label{differentApproach}
\end{align}
which leads back to the renormalizations (\ref{Reno1}). This discussion sheds
another light on what went wrong in the calculation by LI. They overlooked that $Z_{0}$ and
$Z_{1}$ approach their limit $Z$ differently as manifested in Eq.~(\ref{differentApproach}). This difference, when overlooked, leads to erroneous results.

The bare Hamiltonian~(\ref{H_1}) is form-invariant under a rescaling of the
fields that makes one of the coupling constants redundant. This rescaling can
be chosen in particular so that $\mathring{g}_{2}^{\prime}=\mathring{g}_{2}$ (see the discussion after Eqs.~(\ref{Resc1}) and
(\ref{Resc2})) which leads to
the Hamiltonian~(\ref{H_2}) in form of the asymmetric Potts model. Owing to the permutation-symmetry $S_{n}$ of this Hamiltonian, this relation holds
even in renormalized form, $u_{2}^{\prime}=u_{2}$, where $u_{2}$ is related to
the bare $\mathring{g}_{2}$ by the renormalization factor $Z_{2}$. It follows
the relation
\end{subequations}
\begin{equation}
\frac{B_{2}^{\prime}}{u_{2}^{\prime}}=\frac{B_{2}}{u_{2}}=:Z_{2}-1\,,
\label{Zet2}%
\end{equation}
where $Z_{2}$ depends only on scaling invariant combinations of the coupling
constants, say
\begin{subequations}
\label{Inv-Coup}%
\begin{align}
u  &  =u_{2}u_{2}^{\prime}\,,\label{Inv-Coup1}\\
v  &  =u_{1}u_{2}\,,\label{Inv-Coup2}\\
w  &  =u_{0}u_{2}^{3}\,. \label{Inv-Coup3}%
\end{align}

\subsection{Shift-symmetry and Ward identities}

The Hamiltonian~(\ref{H_1}) is, as typical for a $\phi^{3}$-theory,
form-invariant under a shift of the order parameter by an arbitrary constant. To be more specific, the Hamiltonian is form-invariant under
\end{subequations}
\begin{equation}
\varphi\rightarrow\varphi^{\prime}=\varphi+\gamma\,.
\end{equation}
in conjunction with the parameter-change
\begin{subequations}
\begin{align}
\tau &  \rightarrow\tau^{\prime}=\tau+g_{2}\gamma\,,\\
\rho &  \rightarrow\rho^{\prime}=\rho-(2g_{1}+g_{2}^{\prime})\gamma\,,\\
h &  \rightarrow h^{\prime}=h-\tau\gamma-\frac{g_{2}}{2}\gamma^{2}\,.
\end{align}
Note that the coupling-constants are not transformed. Hence, the primed
fields and parameters are renormalized with the same counter-terms as the
original ones. Thus, the transformations represent a scaling symmetry in renormalized as well as in
bare form. We introduce the two-dimensional vector $\underline{f}%
=(-2g_{1}-g_{2}^{\prime},g_{2})=G_{\varepsilon}^{-1/2}\mu^{\varepsilon
/2}\underline{v}$ with $\underline{v}=(-2u_{1}-u_{2}^{\prime},u_{2})$ together
with its bare form $\underline{\mathring{f}}$, define $\mathring{\gamma
}=Z^{1/2}\gamma$, and compare the renormalizations, e.g.,%
\end{subequations}
\begin{align}
Z\mathring{\tau}^{\prime} &  =\underline{\underline{Z}}\cdot\underline
{\tau^{\prime}}=\underline{\underline{Z}}\cdot(\underline{\tau}+\gamma
\underline{f})\nonumber\\
&  =Z(\underline{\mathring{\tau}}+\mathring{\gamma}\underline{\mathring{f}%
})\nonumber\\
&  =\underline{\underline{Z}}\cdot\underline{\tau}+\gamma G_{\varepsilon
}^{-1/2}\mu^{\varepsilon/2}(\underline{v}+\underline{V})\,,
\end{align}
where we have defined $\underline{V}=(-2B_{1}-B_{2}^{\prime},B_{2})$. It
follows the Ward identity%
\begin{equation}
\bigl(\underline{\underline{Z}}-\underline{\underline{1}}\bigr)\cdot
\underline{v}=\underline{V}\,.
\end{equation}
In the same way, we derive a second Ward identity%
\begin{equation}
\bigl(\underline{v}\cdot\underline{\underline{A}}\bigr)_{i}=\delta
_{2,i}-Z_{2,i}\,.
\end{equation}
In particular, we have $B_{2}=-\underline{v}\cdot\underline{\underline{A}}%
\cdot\underline{v}$. Both Ward identities are easily verified at $1$-loop order with the diagrammatic results given in Appendix~\ref{app:1-loopCalc}.
They reduce higher-order calculations enormously,
and lead to important relations between renormalization group functions and
critical exponents. Being linear relations between the counter-terms, the Ward
identities hold for each term of the Laurent-expansions, in particular for the
residua
\begin{subequations}
\label{Ward}%
\begin{align}
&  Z_{2,i}^{(1)}=-\underline{v}\cdot\underline{\underline{A}}^{(1)}%
\,,\label{Ward1}\\
&  \underline{V}^{(1)}=\underline{\underline{Z}}^{(1)}\cdot\underline
{v}\,,\label{Ward2}\\
&  B_{2}^{(1)}=-\underline{v}\cdot\underline{\underline{A}}^{(1)}%
\cdot\underline{v}\,.\label{Ward3}%
\end{align}

It is of some interest to state the Ward identities also in terms of the Vertex functions.
The shift-invariance leads to the following identity for the vertex-function
generating functional (remember that no renormalizations are influenced
by the shift)%
\end{subequations}
\begin{align}
&  \Gamma\lbrack\tilde{\varphi},\varphi;\underline{\tau},h]=\Gamma
\lbrack\tilde{\varphi},\varphi;\underline{\tau}]+(h,\tilde{\varphi
})\nonumber\\
&  \qquad=\Gamma\lbrack\tilde{\varphi},\varphi+\gamma;\underline{\tau
}+\underline{f}\gamma,h-\tau\gamma-\frac{g_{2}}{2}\gamma^{2}]\,.
\label{Vert-Erz}%
\end{align}
Differentiation with respect to $\gamma$ leads to the Ward identities
\begin{equation}
\Gamma_{\tilde{k},k+1}(\{\mathbf{q}=0\})=\tau\delta_{\tilde{k},1}\delta
_{k,0}-\underline{f}\cdot\frac{\partial}{\partial\underline{\tau}}%
\Gamma_{\tilde{k},k}(\{\mathbf{q}=0\})
\end{equation}
between the vertex-functions.

\subsection{RG functions}
\label{subsec:RGfunctions}

RG functions express the change of the renormalized quantities under an
infinitesimal change of the momentum-scale $\mu$ (while holding bare quantities constant). They are the
essential ingredients of the RG equations.
As a scale change between two renormalized and therefore finite theories, the RG functions are themselves finite quantities without $\varepsilon$-poles. We define%
\begin{subequations}
\begin{align}
\beta_{\alpha} &  =\left.  \mu\partial_{\mu}u_{\alpha}\right\vert _{0}%
=-\frac{\varepsilon}{2}u_{\alpha}+\beta_{\alpha}^{(0)}\,,\\
\gamma &  =\left.  \mu\partial_{\mu}\ln Z\right\vert _{0}\,,
\end{align}
where $\beta_{\alpha}^{(0)}$ and $\gamma$ are independent of $\varepsilon$
in minimal renormalization. It follows that
\end{subequations}
\begin{equation}
\left.  \mu\partial_{\mu}\right\vert _{0}(Z,K,\cdots)=-\frac{1}{2}%
u\cdot\partial_{u}(Z^{(1)},K^{(1)},\cdots)+O(\varepsilon^{-1})\, ,
\end{equation}
where we abbreviate $\sum_{\alpha}u_{\alpha}\partial_{u_{\alpha}}%
=:u\cdot\partial_{u}$. Expanding in the following all expressions in
Laurent-series with respect to $\varepsilon$, and making use of the fact that
all renormalized quantities are free of $\varepsilon$-poles, we obtain%
\begin{subequations}
\begin{align}
\beta_{\alpha}^{(0)} &  =\frac{3}{2}\gamma u_{\alpha}-\frac{1}{2}%
\bigl(1-u\cdot\partial_{u}\bigr)B_{\alpha}^{(1)}\,,\\
\gamma &  =-\frac{1}{2}u\cdot\partial_{u}Z^{(1)}\,,\\
\hat{\gamma}^{\prime} &  =-\frac{1}{2}u\cdot\partial_{u}K^{(1)}\,,
\end{align}
so that
\end{subequations}
\begin{subequations}
\begin{align}
\left.  \mu\partial_{\mu}\right\vert _{0}\tilde{\varphi} &  =-\frac{\gamma}%
{2}\tilde{\varphi}\,,\\
\left.  \mu\partial_{\mu}\right\vert _{0}\varphi &  =-\frac{\gamma}{2}%
\varphi-\hat{\gamma}^{\prime}\tilde{\varphi}
\end{align}
in Greens functions. Similarly, we get%
\end{subequations}
\begin{subequations}
\begin{align}
\left.  \mu\partial_{\mu}\right\vert _{0}\underline{\tau} &  =\underline{\tau
}\cdot\underline{\underline{\hat{\kappa}}}\,,\\
\left.  \mu\partial_{\mu}\right\vert _{0}h &  =\frac{\gamma}{2}h+\frac{1}%
{2}G_{\varepsilon}^{1/2}\mu^{-\varepsilon/2}\,(\underline{\tau}\cdot
\underline{\underline{\hat{\alpha}}}\cdot\underline{\tau})\,,
\end{align}
where we have defined%
\end{subequations}
\begin{subequations}
\begin{align}
\underline{\underline{\hat{\kappa}}} &  =\gamma\underline{\underline{1}}%
+\frac{1}{2}u\cdot\partial_{u}\bigl(\underline{\underline{Z}}^{(1)}%
\bigr)^{T}\,,\\
\underline{\underline{\hat{\alpha}}} &  =\frac{1}{2}\bigl(1+u\cdot\partial
_{u}\bigr)\underline{\underline{A}}^{(1)}\,.
\end{align}
It is now easy to derive relations between the Gell-Mann--Low functions with
help of the Ward identities~(\ref{Ward}). We obtain
\end{subequations}
\begin{subequations}
\label{WardGM}%
\begin{align}
\hat{\kappa}_{i,2} &  =\gamma\delta_{i,2}-\bigl(\underline{\underline
{\hat{\alpha}}}\cdot\underline{v}\bigr)_{i}\,,\label{WardGM1}\\
\underline{\hat{\beta}} &  =\frac{\gamma-\varepsilon}{2}\underline
{v}+\underline{v}\cdot\underline{\underline{\hat{\kappa}}}\,,\label{WardGM2}\\
\hat{\beta}_{2} &  =\frac{3\gamma-\varepsilon}{2}u_{2}-\underline{v}%
\cdot\underline{\underline{\hat{\alpha}}}\cdot\underline{v}\,.\label{WardGM3}%
\end{align}
Here we used the two-dimensional vectors $\underline{v}=(-2u_{1}-u_{2}%
^{\prime},u_{2})$ and $\underline{\hat{\beta}}=(-2\beta_{1}-\beta_{2}^{\prime
},\beta_{2})$. In Appendix~\ref{app:1-loopCalc}, we state all the RG functions to $1$-loop
order. With the results given there, the relations (\ref{WardGM}) are verified easily.

\subsection{RG equations}
Now, we derive the RG equations that determine how the quantities featured in our theory transform or flow under variation of the momentum-scale $\mu$. In order for the RG equations to produce reliable results, we have to remove a this stage any remaining scaling-redundancy that could contaminate the RG-flow. For example, if we continued using the variables of Sec.~\ref{subsec:RGfunctions}, we were at risk to erroneously conclude from Eq.~(\ref{WardGM2}) that there is an eigenvalue
$(\varepsilon-\gamma_{\ast})/2$ of the matrix $\underline{\underline
{\hat{\kappa}}}_{\ast}$ at a fixed point $(u_{\alpha})_{\ast}$ with
$\underline{\hat{\beta}}_{\ast}=0$.

To remove the one remaining scaling redundancy from our theory, we switch to rescaling invariant fields
\end{subequations}
\begin{equation}
\phi=u_{2}\varphi,\quad\tilde{\phi}=u_{2}^{-1}\tilde{\varphi}\,,
\end{equation}
control parameters $\underline{t}=(\sigma,\tau)$ with
\begin{subequations}
\begin{align}
\sigma &  =u_{2}^{2}\rho\,,\\
H &  =2g_{2}h\,,
\end{align}
and the dimensionless coupling constants given by Eqs.~(\ref{Inv-Coup}). This procedure yields the new $\beta$-functions
\end{subequations}
\begin{subequations}
\begin{align}
\beta_{u} &  =u_{2}\beta_{2}^{\prime}+u_{2}^{\prime}\beta_{2}\,,\\
\beta_{v} &  =u_{2}\beta_{1}+u_{1}\beta_{2}\,,\\
\beta_{w} &  =u_{2}^{3}\beta_{0}+3u_{0}u_{2}^{2}\beta_{2}\,.
\end{align}
The Gell-Mann--Low functions designated with a hat change to%
\end{subequations}
\begin{subequations}
\begin{align}
\gamma^{\prime} &  =u_{2}^{2}\hat{\gamma}^{\prime}\,,\\
\kappa_{1,1} &  =\hat{\kappa}_{1,1}+\zeta\,,\quad\kappa_{1,2}=u_{2}^{-2}%
\hat{\kappa}_{1,2}\,,\\
\kappa_{2,1} &  =u_{2}^{2}\hat{\kappa}_{2,1}\,,\quad\kappa_{2,2}=\hat{\kappa
}_{2,2}\,,\\
\alpha_{1,1} &  =u_{2}^{-3}\hat{\alpha}_{1,1}\,,\quad\alpha_{1,2}=u_{2}%
^{-1}\hat{\alpha}_{1,2}\,,\quad\alpha_{2,2}=u_{2}\hat{\alpha}_{2,2}\,,
\end{align}
where we have defined
\end{subequations}
\begin{equation}
\zeta=\frac{\beta_{u}}{u}=2\frac{\beta_{2}}{u_{2}}=2\frac{\beta_{2}^{\prime}%
}{u_{2}^{\prime}}\,.
\end{equation}
Note that in case of $u=0$, the function $\zeta$ is in general finite and non-zero.

Now, we are in the position to set up our ultimate RG equations. The generator
$\mathcal{D}_{\mu}$ of the RG, i.e.,  the derivative
$\left.  \mu\partial_{\mu}\right\vert _{0}$ purely expressed in terms of renormalized
parameters, is given by
\begin{equation}
\mathcal{D}_{\mu}=\mu\frac{\partial}{\partial\mu}+\underline{t}\cdot
\underline{\underline{\kappa}}\cdot\frac{\partial}{\partial\underline{t}%
}+\beta_{u}\frac{\partial}{\partial u}+\beta_{v}\frac{\partial}{\partial
v}+\beta_{w}\frac{\partial}{\partial w}\,.\label{RG-Gen}%
\end{equation}
Its application to the fields in a correlation function produces the RG equations
\begin{subequations}
\label{RG-Gl}%
\begin{align}
&  \mathcal{D}_{\mu}\tilde{\phi}=-\frac{\gamma+\zeta}{2}\tilde{\phi
}\,,\nonumber\\
&  \mathcal{D}_{\mu}\phi=-\frac{\gamma-\zeta}{2}\phi-\gamma^{\prime}%
\tilde{\phi}\,.\label{RG-Feld}%
\end{align}
In addition the RG equation of the external field $H$, which is linearly
related to $z$ (the integration variable of the inverse Laplace
transformation) and the control-parameters $\underline{t}$ are%
\end{subequations}
\begin{subequations}
\begin{align}
\mathcal{D}_{\mu}H &  =\frac{\gamma+\zeta+\varepsilon}{2}H+\underline{t}%
\cdot\underline{\underline{\alpha}}\cdot\underline{t}\,,\label{RG-z}\\
\mathcal{D}_{\mu}\underline{t} &  =\underline{t}\cdot\underline{\underline
{\kappa}}\label{RG-cont}%
\end{align}
We introduce the combination%
\end{subequations}
\begin{equation}
a=u+2v=-v_{1}v_{2}\,,
\end{equation}
with the corresponding Gell-Mann--Low function $\beta_{a}=\beta_{u}+2\beta
_{v}$, and the $2$-dimensional orthogonal vectors
\begin{equation}
\underline{w}=(a^{-1},1),\quad\underline{\overline{w}}=(-a,1)\,.
\end{equation}
The Ward identities (\ref{WardGM}) yield
\begin{equation}
\kappa_{i2}=\gamma\delta_{i,2}-(\underline{\overline{w}}\cdot\underline
{\underline{\alpha}})_{i}\,, \label{F-Ward0}%
\end{equation}
and the important relations between RG functions
\begin{subequations}
\label{F-Ward}%
\begin{align}
(\underline{\overline{w}}\cdot\underline{\underline{\kappa}})_{2}  &
=(\varepsilon-\gamma+\zeta)/2\,,\label{F-Ward-1}\\
a^{-1}\beta_{a}  &  =-\underline{\overline{w}}\cdot\underline{\underline
{\kappa}}\cdot\underline{w}\,. \label{F-Ward-2}%
\end{align}
The last equation in combination with the orthogonality of $\underline
{\overline{w}}$ and $\underline{w}$ shows that these vectors are for $\beta_{a}=0$ right and  left eigen-vectors of $\underline{\underline{\kappa}}$, respectively, with eigenvalues
\end{subequations}
\begin{subequations}
\label{Eig-Wert}%
\begin{align}
\kappa_{1}  &  =(\underline{\underline{\kappa}}\cdot\underline{w}%
)_{2}=a(\underline{\underline{\kappa}}\cdot\underline{w})_{1}-a^{-1}\beta
_{a}\,,\label{Eig-Wert1}\\
\kappa_{2}  &  =(\underline{\overline{w}}\cdot\underline{\underline{\kappa}%
})_{2}=(\varepsilon-\gamma+\zeta)/2\,. \label{Eig-Wert2}%
\end{align}
Note that $\kappa_{2}=(\varepsilon+\zeta-\gamma)/2$ determines the RG-flow of
the order-parameter field $\phi$, Eq.\ (\ref{RG-Feld}). This shows that each
control parameter combination proportional to $\underline{\overline{w}}$ is
redundant and can be eliminated by an order-parameter shift. Otherwise, the
combination%
\end{subequations}
\begin{equation}
y:=\underline{t}\cdot\underline{w}=a^{-1}\sigma+\tau\label{KollKontPar}%
\end{equation}
is free of the shift-redundancy and has the the independent scaling exponent
$\kappa_{1}$. We expect that $y$ defines the distance from the collapse
transition line in the phase diagram.

To 1-loop order, our diagrammatic calculation leads to
\begin{subequations}
\label{Beta-Funk}%
\begin{align}
\beta_{u}  &  {=}{\Big(-\varepsilon+}\frac{7}{{2}}u{+}10v\Big)u{\,,}%
\label{Beta-Funk1}\\
\beta_{v}  &  ={\Big(}-\varepsilon+\frac{25}{6}u+\frac{21}{2}v\Big)v-\frac
{5}{6}w\,,\label{Beta-Funk2}\\
\beta_{w}  &  {=\Big(}-2\varepsilon+\frac{21}{2}u+25v\Big)w\nonumber\\
&  -{\Big(}5u^{2}+\frac{29}{2}uv+11v^{2}\Big))v\,,\label{Beta-Funk3}\\
\gamma &  =-\frac{u+4v}{6}\,,\quad\gamma^{\prime}=\frac{2uv+3v^{2}-w}{6}\,,
\label{Gam-Fu}%
\end{align}
and the matrices
\end{subequations}
\begin{subequations}
\label{Matrix}%
\begin{align}
\underline{\underline{\kappa}}  &  =%
\begin{pmatrix}
8(2u+5v)/3-\varepsilon\,, & -1\\
5(w-2uv-3v^{2})/3\,, & 5(u+4v)/6
\end{pmatrix}
\,,\label{Matrix1}\\
\underline{\underline{\alpha}}  &  =%
\begin{pmatrix}
0\,, & 1\\
1\,, & -2v
\end{pmatrix}
\,. \label{Matrix2}%
\end{align}
\end{subequations}
With these 1-loop results, the general results (\ref{F-Ward0}), (\ref{F-Ward} and (\ref{Eig-Wert}), which hold
to all loop-orders, are easily verified.

\subsection{RG flow and fixed points}

\begin{table}[htb]
\par
\begin{center}%
\begin{tabular}
[c]{|c|c|c|c|c|}\hline
& {$u_{\ast}$} & $v_{\ast}$ & $w_{\ast}$ & stability\\\hline
G & $0$ & $0$ & $0$ & $---$\\\hline
C & $0$ & $\frac{\varepsilon\big(69+\sqrt{201}\big)}{760}$ & $\frac
{6\varepsilon^{2}\big(689\sqrt{201}-339\big)}{5\times760^{2}}$ & $+++$\\\hline
P & $\frac{2\varepsilon}{7}$ & $0$ & $0$ & $++-$\\\hline
YL & $-\frac{2\varepsilon}{3}$ & $\varepsilon/3$ & $-\varepsilon^{2}/9$ &
$+--$\\\hline
In1 & $-\frac{\varepsilon}{2}$ & $11\varepsilon/40$ & $-517\varepsilon
^{2}/8000$ & $++-$\\\hline
In2 & $0$ & $\frac{\varepsilon\big(69-\sqrt{201}\big)}{760}$ & $\frac
{-6\varepsilon^{2}\big(689\sqrt{201}+339\big)}{5\times760^{2}}$ &
$+--$\\\hline
\end{tabular}
\end{center}
\caption{\label{tab:fixedPoint}RG fixed points to leading order.}%
\end{table}

\begin{figure}[htb]
\includegraphics[width=7cm]{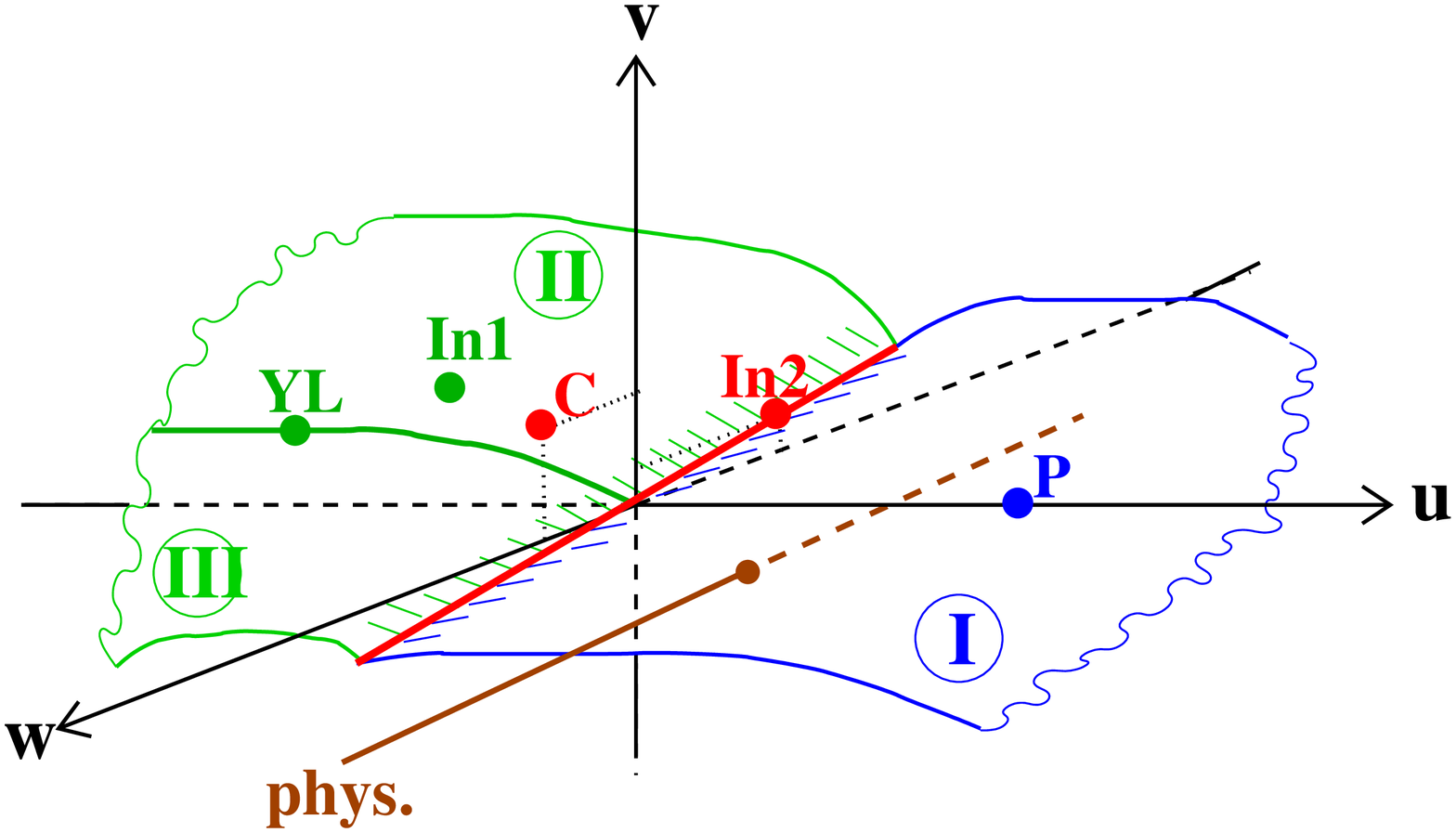}\caption{\label{fig:flow}(Color online)
Sketch of the invariant manifolds of the RG-flow as explained in the main
text.}%
\end{figure}

The fixed points of our RG are determined by the zeros of the  Gell-Mann--Low RG functions for
the three coupling-constants as given
in Eqs.~(\ref{Beta-Funk1}) to (\ref{Beta-Funk3}). The picture of the topology
of the fixed points, invariant lines, and separating surfaces resulting from
the RG flow that arises from these equations in the three-dimensional space
spanned by these coupling-constants is sketched in Fig.~\ref{fig:flow}. The
BRS-plane $u=0$ (red) is an invariant plane of the flow equations
(\ref{Beta-Funk1}) to (\ref{Beta-Funk3}) to all orders and divides the
$(u,v,w)$-space in two parts: the percolation-part with $u>0$ (blue, I) and
the Yang-Lee-part with $u<0$ (green, I and II). The latter part is
non-physical for the branched polymer problem. The percolation line $v=w=0$ is
an invariant line for both signs of $u$. For $u>0$ the flow goes to the
percolation fixed point (P) whereas for $u<0$ the flow tends to infinity. The
Yang-Lee-line (bold green line) with $a=b=0$, where $a=u+2v$ and $b=u^{2}+4w$,
is also an invariant line for both signs of $u$. For $u<0$ the flow goes to
the Yang-Lee fixed point (YL) whereas for $u>0$ the flow runs away to
infinity. Altogether we have six fixed points which are compiled in
Table~\ref{tab:fixedPoint} to $1$-loop order. Besides the trivial Gaussian
fixed point (G) we find in the BRS-plane the stable collapse fixed point (C),
and an instable fixed point (In2). This point lies on a separatrix in the
BRS-plane (bold red line) and is attracting on it. The flow of the part which
contains C is of course attracting to C. The other part shows runaway flow.
Turning to the percolation-part of the $(u,v,w)$-space, there is the
aforementioned instable percolation fixed point P on the percolation line
$v=w=0$. Because P has two stable directions, it defines a separating invariant
surface with P as an attracting fixed point that divides the space in two
parts. The flow in one of it goes to C whereas the flow in the other part is
again running away. The separating surface, the stability plane of P for
$u>0$, is a continuation of the separatrix found above on the BRS-plane for
$u=0$. In the Yang-Lee-part of the $(u,v,w)$-space, we also find a separating
surface which is the continuation of the BRS-separatrix now into the region
with $u<0$. This invariant surface is separated in two parts by the
Yang-Lee-line. One part is attracting to an instable fixed point (In1), the
other part shows runaway flow. Both surfaces divide the $(u,v,w)$-space in a
wedge-shaped part attracting to C, and a part where the flow goes to infinity.
The edge of the wedge is the separatrix in the BRS-plane. Note that the two
separating surfaces are not smoothly connected at the separatrix since the
BRS-plane is itself a separating surface.

The line labelled phys.\ (brown) is closely related to the collapse line in
the phase diagram, Fig.~\ref{fig:phaseDia}, and its meaning is as follows.
Recall that we focus on asymptotically large RBP's, and hence the external
field $h$ is near criticality. The control parameter $y$ and the three coupling
constants are thought to be expressed as functions of the two fugacities
spanning the phase diagram. At the collapse, i.e., when $y$ becomes critical,
the two fugacities are not independent and hence, the coupling-constants can
be parametrized in terms of a single fugacity. Hence the collapse line in the
phase diagram corresponds to a line in the flow diagram which we represent by
the brown line. As long as this line lies above the percolation surface, the
RG flows to C. From the point where the brown line pierces the blue
percolation surface, the RG flows to the percolation fixed point P. From any
point on the line below the percolation surface, the RG runs off to infinity.

Before returning to the $\theta$-transition as our main focus, we would like
point out the following lesson regarding the $\theta^{\prime}$-transition that
our flow diagram teaches. Usually, runaway flows are associated with
fluctuation induced first order transitions. Here, the region below the
percolation surface where the coupling-constants runs away to ever more
positive values indicate that this transition might be discontinuous and not,
as previously assumed, a second order transition.

\subsection{Scaling at the collapse-transition}

Now, we determine the scaling behavior of the order parameter $\langle\tilde{n}\rangle_{z}=\langle
\varphi\rangle_{z}=\Phi/g_{2}$  (here we
have included a factor $g_{2}$ in the definition of $\Phi$ for convenience), the correlation function of $\varphi$ and $\tilde{\varphi}$ and the correlation length.
The external field $H=2g_{2}h$ (which is a
linear function of the Laplace-variable $z$) is related to $\Phi$ via the equation of state%
\begin{equation}
h+\left.  \frac{\delta\Gamma\lbrack\tilde{\varphi},\varphi;\underline{\tau}%
]}{\delta\tilde{\varphi}}\right\vert _{\tilde{\varphi}=0,\varphi=\Phi/g_{2}%
}=0\,, \label{Zust-Gl}%
\end{equation}
where $\Gamma\lbrack\tilde{\varphi},\varphi;\underline{\tau}]$ is the vertex
generating functional. The equation of state guarantees that tadpole
insertions in diagrams are cancelled by the external field $h$, and
$\langle\varphi\rangle=0$ after the shift $\varphi\rightarrow\varphi
+\Phi/g_{2}$. Using again the shift-symmetry of the vertex generating
function, Eq.~(\ref{Vert-Erz}), the equation of state (\ref{Zust-Gl}) is
reduced to%
\begin{equation}
H+\tau^{2}=(\tau-\Phi)^{2}+T(\sigma+a\Phi,\tau-\Phi)\,, \label{H_tadp}%
\end{equation}
where $T(\underline{t})=-2g_{2}\Gamma_{1,0}(\underline{\tau})$ is the sum of
the tadpole-diagrams, which we have calculated to $1$-loop order. To find
$\Phi$ as a function of $z$, we invert equation (\ref{H_tadp}) and obtain
$(\tau-\Phi)$ as a function of $(H+\tau^{2})$ and $y$. The inverse has
according to (\ref{Zust-Gl}) a critical point at a value of $\Phi$ where
\begin{align}
&  g_{2}\frac{\partial}{\partial\Phi}\left(  \left.  \frac{\delta\Gamma
\lbrack\tilde{\varphi},\varphi;\underline{\tau}]}{\delta\tilde{\varphi}%
}\right\vert _{\tilde{\varphi}=0,\varphi=\Phi/g_{2}}\right) \nonumber\\
&  \qquad=\Gamma_{1,1}(\mathbf{q}=0,\sigma+a\Phi,\tau-\Phi) =0\,.
\end{align}
This condition determines eventually the critical value $z_{c}$ of the inverse
Laplace-transformation where the first singularity in the complex $z$-plane is
positioned. It is therefore the value where the correlation length $\xi
(z)\sim1/\sqrt{\Gamma_{1,1}(\mathbf{q}=0)}$ tends to infinity.

To find the scaling behavior of $\Phi$ as a function of $(z-z_{c})$ near this
critical point, we examine the RG flow of the shift-invariant
combinations of control parameters $y=(\tau+a^{-1}\sigma)$, $M=(\tau-\Phi)$,
and $L=(\tau^{2}+H)\sim(z-z_{c})$. Note that at this point the redundant
variable $\tau$ can be set to zero. The RG equations for these combinations
are easily derived from the equations (\ref{RG-Feld}), (\ref{RG-z}) and (\ref{RG-cont})
using the properties which follow from the Ward-identities. They are given by
\begin{subequations}
\label{Fluss}%
\begin{align}
&  \mathcal{D}_{\mu}y=\kappa_{1}y\,,\label{Fluss1}\\
&  \mathcal{D}_{\mu}M=\kappa_{2}M+\kappa_{1,2}ay\,,\label{Fluss2}\\
&  \mathcal{D}_{\mu}L=(\kappa_{2}+\gamma)L+\alpha_{1,1}a^{2}y^{2}\,.
\label{Fluss3}%
\end{align}
The solutions of these flow equations at a fixed point in terms of a flow parameter
$l$ such that $\mu(l)=\mu l$ are given by
\end{subequations}
\begin{subequations}
\label{Int-Fluss}%
\begin{align}
y(l)  &  =l^{\kappa_{1\ast}}y\,,\label{Int-Fluss1}\\
M(l)+p_{1}y(l)  &  =l^{\kappa_{2\ast}}(M+p_{1}y)\,,\label{Int-Fluss2}\\
L(l)+p_{2}y(l)^{2}  &  =l^{(\kappa_{2}+\gamma)_{\ast}}(L+p_{2}y^{2})\,,
\label{Int-Fluss3}%
\end{align}
where $p_{1}=[\kappa_{1,2}a/(\kappa_{2}-\kappa_{1})]_{\ast}$ and
$p_{2}=[\alpha_{1,1}a^{2}/(\kappa_{2}+\gamma-2\kappa_{1})]_{\ast}$. Taking
into account the naive dimensions of $M$, $y$, and $L$, the relation between these
quantities as the inversion of Eq.~(\ref{H_tadp}) is
\end{subequations}
\begin{align}
&  \bigl(M(l)+p_{1}y(l)\bigr)/\mu(l)^{2}\nonumber\\
&  =F(\bigl(L(l)+p_{2}y(l)^{2}\bigr)/\mu(l)^{4},y(l)/\mu(l)^{2})
\end{align}
 in dimensionless form.
Choosing $l$ so that $\bigl(L(l)+p_{2}y(l)^{2}\bigr)/\mu
(l)^{4}=1$, we obtain the order-parameter equation in scaling form
\begin{align}
M+p_{1}y  &  =\bigl(L+p_{2}y^{2}\bigr)^{\beta/\Delta}\nonumber\\
&  \times\mathcal{F}\left(  y/\bigl(L+p_{2}y^{2}\bigr)^{1/\Delta}\right)  \,,
\end{align}
and setting $(L+p_{2}y^{2})\sim(z-z_{c})$ and $(M+p_{1}y)\sim(\Phi_{c}-\Phi)$, we obtain
\begin{equation}
\Phi_{c}-\Phi=(z-z_{c})^{\beta/\Delta}\mathcal{F}_{\Phi}\left(  y/(z-z_{c}%
)^{1/\Delta}\right)  \,. \label{Skal-Zust-Gl}%
\end{equation}
Here, the scaling function $\mathcal{F}_{\Phi}$ is identical to $\mathcal{F}$
up to some non-interesting constant factors, and the critical exponents are
given by the fixed point values of the various RG-functions
\begin{subequations}
\label{kritExp}%
\begin{align}
1/\nu &  =2-\kappa{_{1\ast}\,,\quad}\eta=\gamma_{\ast}-\zeta_{\ast}%
\,,\quad\tilde{\eta}=\gamma_{\ast}+\zeta_{\ast}\,,\label{kritExp1}\\
\beta/\nu &  =2-\kappa{_{2\ast}=}\frac{1}{2}(d-2+\eta{)\,,}\label{kritExp2}\\
\Delta/\nu &  =4-\kappa{_{2\ast}-}\gamma_{\ast}=\frac{1}{2}(d+2-\tilde{\eta
}{)\,.} \label{kritExp3}%
\end{align}
If $\zeta_{\ast}\neq0$, which happens if $u_{\ast}=0$ and thus holds true at
the collapse-transition, we find three independent critical exponents $\eta$,
$\tilde{\eta}$, and $\nu$.

The RG equation for the correlation function $G_{1,1}(\mathbf{r})=\langle
\phi(\mathbf{r})\tilde{\phi}(\mathbf{0})\rangle_{z}^{(cum)}$  follows from
Eq.~(\ref{RG-Feld}) as%
\end{subequations}
\begin{equation}
\Big(\mathcal{D}_{\mu}+\gamma_{\ast}\Big)G_{1,1}(\mathbf{r})=0\
\end{equation}
at a fixed point. Using again the flow parameter $l$, we obtain the solution%
\begin{align}
&  G_{1,1}(\mathbf{r},y,M+p_{1}y,\mu)\nonumber\\
&  =l^{\gamma_{\ast}}G_{1,1}(\mathbf{r},l^{\kappa_{1\ast}}y,l^{\kappa_{2\ast}%
}(M+p_{1}y),\mu l)\nonumber\\
&  =l^{d-2+\gamma_{\ast}}G_{1,1}(l\mathbf{r},y/l^{1/\nu},(M+p_{1}%
y)/l^{\beta/\nu},\mu)\,.
\end{align}
Taking $y$ and $(z-z_{c})$ as independent variables, and expressing
$(M+p_{1}y)$ through the equation of state (\ref{Skal-Zust-Gl}), we find after
choosing $l$  as above the scaling form%
\begin{equation}
G_{1,1}(\mathbf{r},z)=\frac{\mathcal{G}_{1,1}(\mathbf{r}(z-z_{c})^{\nu/\Delta
},y/(z-z_{c})^{1/\Delta})}{\left\vert \mathbf{r}\right\vert ^{d-2+(\eta
+\tilde{\eta})/2}}\,.
\end{equation}
The correlation length $\xi$ is defined by%
\begin{align}
\xi^{2}  &  =\frac{1}{2d}\int d^{d}r\,\mathbf{r}^{2}G_{1,1}(\mathbf{r})/\int
d^{d}r\,G_{1,1}(\mathbf{r})\nonumber\\
&  =\left.  \frac{\partial\ln\Gamma_{1,1}(\mathbf{q})}{\partial q^{2}%
}\right\vert _{\mathbf{q}=0}\,, \label{KorrL-Def}%
\end{align}
where the vertex function $\Gamma_{1,1}(\mathbf{r})$ is related to the
Fourier-transformed correlation function by $\tilde{G}_{1,1}(\mathbf{q}%
)=1/\Gamma_{1,1}(\mathbf{q})$. Hence the correlation length scales as%
\begin{equation}
\xi(z)\sim(z-z_{c})^{-\nu/\Delta}\,. \label{KorrL}%
\end{equation}
In terms of $\xi$, the correlation function is given by%
\begin{equation}
G_{1,1}(\mathbf{r},z)=\frac{\mathcal{G}_{1,1}(\mathbf{r}/\xi,y\xi^{1/\nu}%
)}{\left\vert \mathbf{r}\right\vert ^{d-2+(\eta+\tilde{\eta})/2}}\,.
\label{KorrFu}%
\end{equation}

\section{Observables of the collapsing branched polymer}
\label{sec:RBPobservables}
In this section, we translate our RG results into a language that is more geared towards polymer physics. In particular, we extract the probability distribution $\mathcal{P} (N)$, the radius of gyration and the shape function. As explained in detail in Sec.~\ref{sec:modelling}, these kind of quantities as functions of $N$ are related to the quantities native to our field theory via inverse Laplace transformation.

\subsection{Scaling behavior}

The probability distribution $\mathcal{P} (N)$ is given by Eq.~(\ref{Probab_N}), and asymptotically for $N\gg1$,
we derive%
\begin{align}
\mathcal{P}(N)  &  \sim\mathrm{e}^{z_{c}N}\int_{0}^{\infty}dx\,\frac
{\operatorname*{Disc}\Phi(z_{c}-x)}{2\pi i}\mathrm{e}^{-xN}\nonumber\\
&  \sim\mathrm{e}^{z_{c}N}\int_{0}^{\infty}dx\,\frac{\operatorname*{Disc}%
\Big[(-x)^{\beta/\Delta}\mathcal{F}_{\Phi}\left(  y/(-x)^{1/\Delta}\right)
\Big]}{2\pi i}\mathrm{e}^{-xN}\nonumber\\
&  \sim N^{-1-\beta/\Delta}\mathrm{e}^{z_{c}N}\nonumber\\
&  \times\int_{0}^{\infty}dx^{\prime}\,\frac{\operatorname*{Disc}%
\Big[(-x)^{\beta/\Delta}\mathcal{F}_{\Phi}\left(  N^{1/\Delta}y/(-x^{\prime
})^{1/\Delta}\right)  \Big]}{2\pi i}\mathrm{e}^{-x^{\prime}}\,.
\end{align}
Hence, we immediately obtain the asymptotic scaling form of the animal numbers
from Eq.~(\ref{A_zu_P}) as%
\begin{equation}
\mathcal{A}(N)\sim N^{-1}\mathcal{P}(N)\sim N^{-\theta}\kappa^{N}%
f_{\mathcal{A}}(yN^{\phi})\,, \label{Skal-An}%
\end{equation}
where the animal exponent $\theta$ and the crossover exponent $\phi$ are given
by
\begin{subequations}
\label{PolExp}%
\begin{align}
\theta &  =2+\beta/\Delta=2+\frac{d-2+\eta}{d+2-\tilde{\eta}}%
\,,\label{PolExp1}\\
\phi &  =\frac{1}{\Delta}=\frac{2}{\nu(d+2-\tilde{\eta})}\,. \label{PolExp2}%
\end{align}

In the same way we find the scaling behavior of the monomer-distribution of a
collapsing branched polymer which was calculated in mean-field theory in
Eq.~(\ref{mf-KorrF}). Here we derive from the correlation function that
(\ref{KorrFu})%
\end{subequations}
\begin{align}
G_{N}(\mathbf{r})  &  =\frac{1}{\mathcal{P}(N)}\int_{\sigma-i\infty}%
^{\sigma+i\infty}\frac{dz}{2\pi i}\,\mathrm{e}^{zN}G_{1,1}(\mathbf{r}%
;z)\nonumber\\
&  =\frac{N^{\theta-1}}{\left\vert \mathbf{r}\right\vert ^{d-2+(\eta
+\tilde{\eta})/2}}\nonumber\\
&  \times\int_{0}^{\infty}dx\,\frac{\operatorname*{Disc}\mathcal{G}%
_{1,1}(\mathbf{r}(-x)^{\nu/\Delta},y/(-x)^{1/\Delta})}{2\pi i}\mathrm{e}%
^{-xN}\nonumber\\
&  =\frac{N^{\theta-1}}{\left\vert \mathbf{r}\right\vert ^{d-2+(\eta
+\tilde{\eta})/2}}G(\mathbf{r}/N^{\nu/\Delta},yN^{1/\Delta})\,.
\end{align}
Defining the radius of gyration $R_{N}$ as in Eq.~(\ref{mf-GyrRad}), we write the
monomer distribution in the scaling form%
\begin{equation}
G_{N}(\mathbf{r})=\frac{N}{R_{N}^{d}}\mathcal{G}(\left\vert \mathbf{r}%
\right\vert /R_{N},yN^{\phi}) \label{mon-distr}%
\end{equation}
with the radius of gyration
\begin{equation}
R_{N}=N^{\nu_{A}}\mathcal{R}(yN^{\phi})\,. \label{GyrRad}%
\end{equation}
Its exponent is given by
\begin{equation}
\nu_{A}=\nu/\Delta=\frac{2}{d+2-\tilde{\eta}}\,. \label{GyrExp}%
\end{equation}
As it should, our result satisfies the sum rule%
\begin{equation}
\int d^{d}x\, \mathcal{G}(\mathbf{x},yN^{\phi})=1
\end{equation}

Next, we state our $\varepsilon$-expansion results for the exponents
governing the collapse transition. Thus far, when it came to the diagrammatic part of our theory, we centered our discussion around the 1-loop order of our calculation to keep matters as simple as possible. Our actual calculation, however, went to higher order which allows us to present here results for the critical exponents of the $\theta$-transition to second order in $\varepsilon$. Details of this calculation will be presented elsewhere~\cite{JaSt-prep}. For completeness, we list in Appendix~\ref{app:2-loopResults} our 2-loop results for the RG functions that went into the calculation of the critical exponents.
For the three independent exponents defined
in Eqs.~(\ref{PolExp}) and (\ref{GyrExp}), we obtain
\begin{subequations}
\label{ExpPol}%
\begin{align}
\theta=  &  \frac{5}{2}-0.4925\,(\varepsilon/6)-0.5778\,(\varepsilon
/6)^{2}{\,,}\label{ExpPol1}\\
\phi=  &  \frac{1}{2}+0.0225\,({\varepsilon/6})-0.3580\,({\varepsilon/6}%
)^{2}\,,\label{ExpPol2}\\
\nu_{A}=  &  \frac{1}{4}+0.1915\,(\varepsilon/6)+0.0841\,(\varepsilon
/6)^{2}{\,,} \label{ExpPol3}%
\end{align}
From these expansions, we derive numerical results of the exponents for dimensions $2$ to $5$ by performing simple Pad\'{e}-estimates~\cite{Am84, ZJ02}. These results are compiled in Table~\ref{tab:exponents}.
\begin{table}[htb]
\par
\begin{center}%
\begin{tabular}
[c]{|c|ccc|}\hline\hline
$d$ & $\quad\theta$ & $\quad\phi$ & $\quad\nu_{\mathrm{A}}$\\\hline
$2$ & $1.96(4)$ & $0.37(2)$ & $0.52(3)$\\
$3$ & $2.13(2)$ & $0.427(5)$ & $0.396(7)$\\
$4$ & $2.277(5)$ & $0.469(1)$ & $0.329(2)$\\
$5$ & $2.4025(6)$ & $0.49383(2)$ & $0.2849(2)$\\
$6$ & $2.5$ & $0.5$ & $0.25$\\\hline\hline
\end{tabular}%
\end{center}
\caption{\label{tab:exponents}Pad\'{e}-estimates of the critical exponents.}%
\end{table}

For $d=2$ dimensions, there exist numerical results to which our
$\varepsilon$-expansion results can be compared. Simulations by Hsu and
Grassberger~\cite{HsGr05} for the tree-part of the collapse transition produce
$\theta=1.845$ and $\nu_{A}=0.5362$. These results compares partially satisfactory within the expectations for such a big value of $\varepsilon$. To improve the agreement between our theoretical predictions and the simulations or potential experiments, it is desirable to extent our calculation to higher order \cite{JaSt-prep2}, and apply more sophisticated resummation methods.

Next, we consider corrections to scaling. To determine the leading corrections, it is useful to distinguish between two phenomena. First, there is the irrelevance of cycles near the $\theta$-transition and the associated crossover to tree-behavior with BRS symmetry. For this crossover, the coupling constant $u$ is proportional to the cycle fugacity $z_{cy}$. Using the RG result $u(l)=ul^{\zeta_*}$ and choosing a small parameter $l$ proportional to $R^{-1}_N$, we find that this crossover leads to a correction to all scaling functions proportional to $u/N^{x_u}$, where
\begin{equation}
x_u = \nu_A \zeta_* = d\nu_A +1-\Theta
\end{equation}
is the corresponding crossover exponent. Second, there is the approach of the coupling constants $v$ and $w$ to their fixed-point values.
This approach is described by the eigenvalues of the matrix of first derivatives of the functions $\beta_v$ and $\beta_w$, respectively,
\begin{equation}
\omega_1 = \varepsilon-0.7614\,\varepsilon^2\,,\quad \omega_2 = 1.0344\,\varepsilon-0.6830\,\varepsilon^2\,.
\end{equation}
These so-called Wegner exponents lead to corrections proportional to $N^{-x_i}$ with  $x_i=\nu_A\omega_i$.

\subsection{The shape of the collapsing branched polymer}

Here we will derive the asymptotic forms of the shape function $\mathcal{G}%
(\mathbf{r}/R_{N},yN^{\phi})$, Eq.~(\ref{mon-distr}), of the monomer
distribution for small and large $\left\vert \mathbf{r}\right\vert /R_{N}$ at
the collapse transition line $y=0$. We use methods analogous to methods
applied in \cite{dCl74,Fi66,McKMo71} to the case of linear polymers.

In a first and somewhat hand-waving approach, we assume that the monomer distribution in the interior of
the branched polymer is independent of the size $N$. Hence, for $x\rightarrow
0$, we should have
\end{subequations}
\begin{equation}
\mathcal{G}(x,0)\sim x^{-d+1/\nu_{A}}\,,
\end{equation}
leading to the monomer distribution for $\left\vert \mathbf{r}\right\vert \ll
R_{N}$
\begin{equation}
G_{N}(\mathbf{r})\sim\frac{1}{\left\vert \mathbf{r}\right\vert ^{d-1/\nu_{A}}%
}\,. \label{mon-distr-small}%
\end{equation}
Next, we derive this result more rigorously by application of the short distance
expansion. The leading terms of the operator product
expansions are given by
\begin{subequations}
\label{OpPr}%
\begin{align}
\tilde{\phi}(\mathbf{r}+\mathbf{x}/2)\tilde{\phi}(\mathbf{r}-\mathbf{x}/2)  &
=c_{1}(\mathbf{x},\mu)\tilde{\phi}(\mathbf{r})\,,\label{OpPr1}\\
\phi(\mathbf{r}+\mathbf{x}/2)\tilde{\phi}(\mathbf{r}-\mathbf{x}/2)  &
=c_{2}(\mathbf{x},\mu)\tilde{\phi}(\mathbf{r})+c_{3}(\mathbf{x},\mu
)\phi(\mathbf{r})\,,\label{OpPr2}\\
\phi(\mathbf{r}+\mathbf{x}/2)\phi(\mathbf{r}-\mathbf{x}/2)  &  =c_{4}%
(\mathbf{x},\mu)\tilde{\phi}(\mathbf{r})+c_{5}(\mathbf{x},\mu)\phi
(\mathbf{r})\,. \label{OpPr3}%
\end{align}
The form of these expansions is dictated by the symmetry of our model:
$\tilde{\varphi}$ belongs to the trivial representation of the permutation
group $S_{n\rightarrow0}$, and $\varphi$ has components belonging to the
trivial and the fundamental representation. The scaling behavior of the
functions $c_{i}(\mathbf{x},\mu)\sim\mu^{(d-2)/2}$ follows from the RGE.
Applying the RG differential $\mathcal{D}_{\mu}$ operator to both sides of
(\ref{OpPr}) and comparing the results, we find
\end{subequations}
\begin{equation}
\mathcal{D}_{\mu}c_{1,3}(\mathbf{x},\mu)=-\frac{\tilde{\eta}}{2}%
c_{1,3}(\mathbf{x},\mu)\
\end{equation}
at the collapse fixed point. Hence
\begin{align}
&  c_{1,3}(\mathbf{x},\mu)=l^{\tilde{\eta}/2}c_{1,3}(\mathbf{x},l\mu
)\nonumber\\
&  =(l\mu)^{(d-2)/2}l^{\tilde{\eta}/2}c_{1,3}(l\mu\mathbf{x},1)=\frac
{c_{1,3}(1,1)}{\mu^{\tilde{\eta}/2}\left\vert \mathbf{x}\right\vert
^{d-1/\nu_{A}}}\,.
\end{align}
Using Eq.~(\ref{OpPr2}), we obtain%
\begin{equation}
G_{1,1}(\mathbf{r};z)\sim\frac{\Phi(z)}{\left\vert \mathbf{r}\right\vert
^{d-1/\nu_{A}}}\,. \label{SlopArg}%
\end{equation}
This argument has to be taken with a grain of salt. Strictly speaking, the operator product expansion has to
be inserted in Greens functions that are superficially convergent, otherwise
one has to deal with additive renormalizations. Therefore $G_{1,1}%
(\mathbf{r};z)$ in Eq.~(\ref{SlopArg}) is determined only up to a polynomial in $z$.
However, this polynomial is cancelled by the inverse Laplace transformation as
long as $N>0$. Hence, after the application of the inverse Laplace
transformation to Eq.~(\ref{SlopArg}) and division by $\mathcal{P}(N)$, we indeed get
the result stated in Eq.~(\ref{mon-distr-small}).

Now we turn to the large $\left\vert \mathbf{r}\right\vert $ (or small
$\left\vert \mathbf{q}\right\vert $) behavior of the correlation function. In
this regime, the appropriate vertex function is well approximated by%
\begin{subequations}
\begin{align}
\Gamma_{1,1}(\mathbf{q},z)  &  \approx\Gamma_{1,1}(\mathbf{0},z)\Big(1+\xi
(z)^{2}\mathbf{q}^{2}\Big)\,,\\
\Gamma_{1,1}(\mathbf{0},z)  &  \sim\xi(z)^{-2+(\eta+\tilde{\eta})/2}\,.
\end{align}
The correlation function has the representation%
\end{subequations}
\begin{equation}
G_{1,1}(\mathbf{r};z)\sim\xi(z)^{-(\eta+\tilde{\eta})/2}\int_{0}^{\infty
}ds\,\exp\Big(-\xi(z)^{-2}s-\mathbf{r}^{2}/4s\Big)\,.
\end{equation}
Taking the conditions $\mathbf{r}^{2}/\xi(z)^{2}\gg1$, $N\gg1$ into
consideration, we calculate the monomer distribution employing a double saddle-point
approximation of the $s$- and $z$-integral. We find the distribution in the
form of Eq.~(\ref{mon-distr}) with the shape function%
\begin{equation}
\mathcal{G}(x,0)\sim x^{-t}\exp\Big(-cx^{1/(1-\nu_{A})}\Big)\,.
\end{equation}
$c$ is a constant, and the exponent is%
\begin{equation}
t=d-\frac{d/2-2+\theta}{1-\nu_{A}}\,.
\end{equation}

\subsection{Fractal dimensions}
We conclude this section by briefly discussing the fractal dimensions associated with RBPs. As discussed on several occasions in this paper, collapsing RBPs have a tree-like structure, i.e., they are quasi one dimensional. Thus, the dimension $d_{min}$ of the shortest path between two points on the polymer, also known as the chemical distance, the backbone dimension $d_{bb}$ and the resistor dimension $d_{rr}$ coincide. The fractal dimension $d_f$ governing the total mass of the RBP is $d_f=1/\nu_A$, and the exponent for random walks on a RPB is given by $d_w=d_{min}+d_f$. From what we have presented thus far in this paper, we know $d_f$ to 2-loop order. Knowing the other fractal dimensions requires to calculate $d_{min}$, which is identical to the dynamical exponent $z$ of our model. This calculation is beyond the scope of this paper and will be presented elsewhere~ \cite{JaSt-prep}. For completeness, however,  we find it useful to mention here the results of our dynamical calculation. For the $\theta$-transition, we find
\begin{equation}
d_{min} = 2 - 0.8756\, (\varepsilon/6) - 1.1528\, (\varepsilon/6)^2\,.
\end{equation}
For the swollen RPBs, we obtain
\begin{equation}
d_{min} = 2 - (\varepsilon/9) - \frac{35}{18}\, (\varepsilon/9)^2\,,
\end{equation}
where $\varepsilon=8-d$ because $d=8$ is the upper critical dimension for the swollen phase.=Pad\'{e}-estimates are given in Table \ref{tab:dmin}.
\begin{table}[htb]
\par
\begin{center}
\begin{tabular}
[c]{|c|cc|}\hline\hline
$d$ & $d_{\mathrm{\min}} \, \mathrm{(swollen)}$ & $d_{\mathrm{\min}%
} \, \mathrm{(collaps)}$\\\hline
$2$ & $1.09$ & $1,21$\\
$3$ & $1.22$ & $1.415$\\
$4$ & $1.37$ & $1.624$\\
$5$ & $1.536$ & $1.8277$\\
$6$ & $1.707$ & $2$\\
$7$ & $1.868$ & $2$\\
$8$ & $2$ & $2$\\ \hline\hline
\end{tabular}
\end{center}
\caption{\label{tab:dmin}Pad\'{e}-estimates of the minimal dimension.}
\end{table}%

\section{Concluding remarks and outlook}

\label{sec:concludingRemarks} In summary, we developed a new, dynamical field
theory for isotropic randomly branched polymers, and we used this model in
conjunction with the RG to take a fresh look at this classical problem of
statistical physics. We demonstrated that our model provides an alternative
vantage point to understand the swollen phase via dimensional reduction. We
corrected and pushed ahead the critical exponents for the $\theta$-transition.
We showed that at the stable fixed point the model has BRS symmetry. Hence, asymptotically the RBPs are dominated by tree configurations.
Our RG analysis produces evidence for the $\theta^{\prime}$-transition being a
fluctuation induced first order transition and not as previously assumed a
second order transition. It would be interesting to see if future experimental
or numerical studies can confirm the latter finding.

Complementary to the quasi-static RG analysis presented in this paper, we have also conducted a field theoretic calculation of the dynamical exponent $z$ of our dynamical model~\cite{JaSt-prep}. This calculation produced the first-ever field theoretic results, quoted above, for the fractal dimension $d_{min}$ of the shortest path and related fractal dimensions for RBPs. We are currently completing a three-loop calculation of the asymmetric Potts-model. This calculation pushes the exponents $\theta$, $\phi$ and $\nu_A$ to third order in $\varepsilon$~\cite{JaSt-prep2}.

\appendix

\section{The quasi-static limit}
\label{app:quasi-staticLimit}

This Appendix provides some background on the quasi-static limit that we invoke in Sec.~\ref{sec:modelling} in the derivation of our field theoretic Hamiltonian. Let us consider a dynamic response functional of the general form
\begin{equation}
\mathcal{J}[\tilde{n},n]=\int d^{d}xdt\lambda\tilde{n}\Big[\lambda
^{-1}\partial_{t}+\tau-\nabla^{2}\Big]n+\mathcal{W}[\tilde{n},n]\,,
\end{equation}
where the interaction-part $\mathcal{W}$ reduces to a time-independent
functional $\overline{\mathcal{W}}[\tilde{n}_{0},m_{\infty}]$ of $\tilde
{n}_{0}(\mathbf{r})$ and $m_{\infty}(\mathbf{r})=\lambda\int_{-\infty
}^{+\infty}dtn(\mathbf{r},t)$ after setting $\tilde{n}(\mathbf{r}%
,t)\rightarrow\tilde{n}_{0}(\mathbf{r})=\tilde{n}(\mathbf{r},0)$. We
define%
\begin{align}
\mathcal{H}_{qs}[\tilde{n}_{0},m_{\infty}]  &  :=\mathcal{J}[\tilde{n}%
_{0},n]\nonumber\\
&  =\int d^{d}x\tilde{n}_{0}\Big[\tau-\nabla^{2}\Big]m_{\infty}+\overline
{\mathcal{W}}[\tilde{n}_{0},m_{\infty}]\,,
\end{align}
where $\mathcal{H}_{qs}$ denotes the quasi-static Hamiltonian. The free causal
propagator
\begin{equation}
G(\mathbf{r}-\mathbf{r}^{\prime},t-t^{\prime})=\langle n(\mathbf{r}%
,t)\tilde{n}(\mathbf{r}^{\prime},t^{\prime})\rangle_{0}\sim\theta(t-t^{\prime
})
\end{equation}
with $\theta(t)=1$ if $t>0$ and $\theta(t)=0$ if $t\leq0$ becomes the static
propagator of $\mathcal{H}_{qs}$ after time integration%
\begin{align}
&  \lambda\int_{-\infty}^{\infty}dt\langle n(\mathbf{r},t)\tilde{n}%
(\mathbf{r}^{\prime},t^{\prime})\rangle_{0}=\lambda\int_{0}^{\infty}dt\langle
n(\mathbf{r},t)\tilde{n}(\mathbf{r}^{\prime},0)\rangle_{0}\nonumber\\
&  \qquad\qquad=\langle m_{\infty}(\mathbf{r})\tilde{n}_{0}(\mathbf{r}%
^{\prime})\rangle_{0}=G_{st}(\mathbf{r}-\mathbf{r}^{\prime})\,.
\end{align}
Now consider a diagram of the graphical perturbation expansion of the
connected correlation function $\langle\prod_{i}m_{\infty}(\mathbf{r}%
_{i})\prod_{j}\tilde{n}(\mathbf{r}_{j},0)\rangle$. By causality, the vertices
of the diagram are ordered in time from `left' (i.e., the largest time
involved) to `right' (the smallest time), $\tilde{n}$-legs are left-going,
$n$-legs are right-going. Consider the first vertex which has only propagators
(we suppress the space arguments) $\langle m_{\infty}\tilde{n}(t_{1}%
)\rangle_{0}=\langle m_{\infty}\tilde{n}_{0}\rangle_{0}$ on its $\tilde{n}%
$-legs. Hence, the time-dependence of the $\tilde{n}$-legs of this vertex is
absorbed by the $m_{\infty}$, each $\tilde{n}(t_{1})$ becomes a
time-independent $\tilde{n}_{0}$, and after integration over the vertex-time
$t_{1}$, the integrated vertex becomes a vertex generated by the quasi-static
interaction $\overline{\mathcal{W}}[\tilde{n}_{0},m_{\infty}]$. By induction,
one can prove that this mechanism carries through all the way to and including the last vertex.
 The full diagram
is therefore generated only by static propagators and the interaction-vertices
of the quasi-static Hamiltonian $\mathcal{H}_{qs}[\tilde{n}_{0},m_{\infty}]$.
By itself, however, the quasi-static Hamiltonian is insufficient to describe
the static properties of the theory. As a remnant of its dynamical origin,
$\mathcal{H}_{qs}$ must be supplemented with the causality rule that forbids
the former time-closed propagator loops. Hence the terminology {\em quasi-}static.

\section{1-loop perturbation theory}
\label{app:1-loopCalc}

\begin{figure}[ptb]
\label{fig:corrdias} \includegraphics[width=6cm]{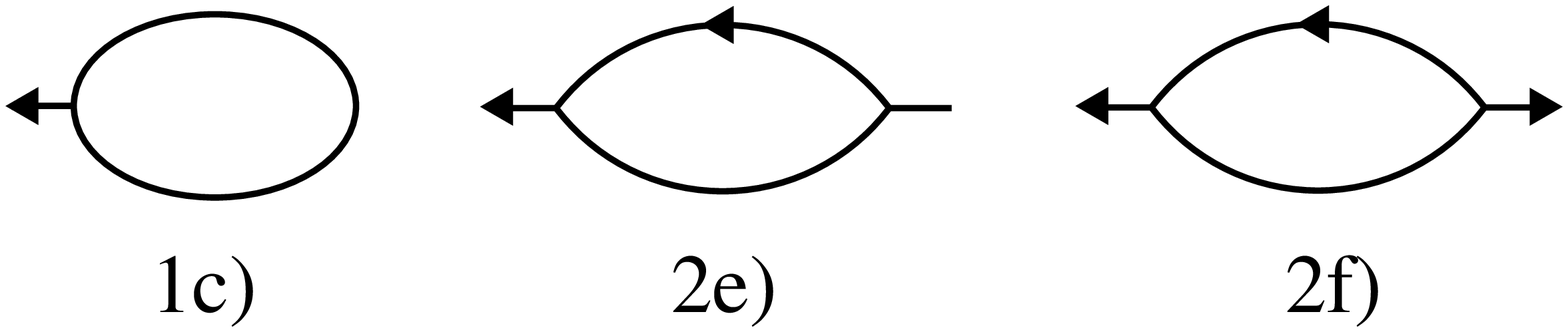}\caption{1-loop
diagrams with a correlator.}%
\end{figure}

In this Appendix we assemble and list our results for the superficially diverging vertex functions $\Gamma_{1,0}$, $\Gamma_{1,1}$,
$\Gamma_{2,0}$, $\Gamma_{1,2}$, $\Gamma_{2,1}$, and $\Gamma_{3,0}$ in the case
$\rho=0$. Recall that we have already calculated the decorations of the diagrams contributing to these vertex functions in Sec.~\ref{sec:modelling}. Thus it remains to perform the integrations over the internal momenta of these diagrams. There are three types of integrals appearing:
\begin{align}
&  I_{1}(\tau)=\int_{\mathbf{p}}\frac{1}{\tau+\mathbf{p}^{2}}=\frac
{G_{\varepsilon}\tau^{2-\varepsilon/2}}{(1-\varepsilon/4)(1-\varepsilon
/2)\varepsilon}\,,\\
&  I_{2}(\tau,\mathbf{q})=\int_{\mathbf{p}}\frac{1}{\bigl(\tau+\mathbf{p}%
^{2}\bigr)\bigl(\tau+(\mathbf{p}+\mathbf{q})^{2}\bigr)}\nonumber\\
&  \qquad=-\frac{2G_{\varepsilon}\tau^{1-\varepsilon/2}}{(1-\varepsilon
/2)\varepsilon}-\frac{(1-\varepsilon/4)G_{\varepsilon}\tau^{-\varepsilon/2}%
}{3(1-\varepsilon/6)\varepsilon}\mathbf{q}^{2}\,,\\
&  I_{3}(\tau)=\int_{\mathbf{p}}\frac{1}{\bigl(\tau+\mathbf{p}^{2}\bigr)^{3}%
}=\frac{G_{\varepsilon}\tau^{-\varepsilon/2}}{\varepsilon}\,,
\end{align}
where we have dropped the UV convergent parts of the integrals which are unimportant for our purposes. In addition to the ($\rho=0$)-diagrams listed in Sec.~\ref{sec:modelling}, we need a few more diagrams that determine the renormalization of $\rho$. Those are the diagrams with an insertion of a $\rho$-vertex, or in other words, diagrams where a propagator is replaced by a correlator, see Fig.~(\ref{fig:corrdias}). These diagrams can be expressed as
\begin{align}
&  1c)=-\frac{g_{2}}{2}\rho I_{2}(\tau,\mathbf{0})\,,\\
&  2e)=-g_{2}^{2}\rho I_{3}(\tau)\,,\\
&  2f)=2g_{2}(2g_{1}+g_{2}^{\prime})\rho I_{3}(\tau)\,.
\end{align}
Altogether we obtain the $\varepsilon$-pole contributions%
\begin{align}
&  \Gamma_{1,0}=h-(1b)-(1c)\nonumber\\
&  \qquad=h+\frac{G_{\varepsilon}\tau^{-\varepsilon/2}}{\varepsilon
}\bigl(g_{1}\tau^{2}-g_{2}\tau\rho\bigr)\,,\\
&  \Gamma_{1,1}=(\tau+\mathbf{q}^{2})-(2b)-(2e)\nonumber\\
&  \qquad=\Big\{\tau-\frac{G_{\varepsilon}\tau^{-\varepsilon/2}}{\varepsilon
}\bigl[g_{2}\bigl(4g_{1}+g_{2}^{\prime}\bigr)\tau-g_{2}^{2}\rho
\bigr]\Big\}\nonumber\\
&  \qquad\qquad+\Big\{1-\frac{G_{\varepsilon}\tau^{-\varepsilon/2}%
}{6\varepsilon}g_{2}\bigl(4g_{1}+g_{2}^{\prime}\bigr)\Big\}\mathbf{q}^{2}\,,\\
&  \Gamma_{2,0}=\rho-(2c)-(2f)\nonumber\\
&  \qquad=\Big\{\rho-2\frac{G_{\varepsilon}\tau^{-\varepsilon/2}}{\varepsilon
}\bigl[\bigl(g_{0}g_{2}-3g_{1}^{2}-2g_{1}g_{2}^{\prime}\bigr)\tau\nonumber\\
&  \qquad\qquad\qquad\qquad\qquad\qquad+\bigl(2g_{1}g_{2}+g_{2}g_{2}^{\prime
}\bigr)\rho\bigr]\Big\}\nonumber\\
&  \qquad+\Big\{1-\frac{G_{\varepsilon}\tau^{-\varepsilon/2}}{3\varepsilon
}\bigl(g_{0}g_{2}-3g_{1}^{2}-2g_{1}g_{2}^{\prime}\bigr)\Big\}\mathbf{q}%
^{2}\,,\\
&  \Gamma_{1,2}=-g_{2}-(3b)\nonumber\\
&  \qquad=-\bigl[1-2\frac{G_{\varepsilon}\tau^{-\varepsilon/2}}{\varepsilon
}\bigl(3g_{1}g_{2}+g_{2}g_{2}^{\prime}\bigr)\bigr]g_{2}\,,\\
&  \Gamma_{2,1}=\bigl(2g_{1}+g_{2}^{\prime}\bigr)-(3c)\nonumber\\
&  \qquad=2g_{1}-2\frac{G_{\varepsilon}\tau^{-\varepsilon/2}}{\varepsilon
}\bigl[\bigl(7g_{1}g_{2}+3g_{2}g_{2}^{\prime}\bigr)g_{1}-g_{2}g_{0}%
\bigr]\nonumber\\
&  \qquad\qquad+\bigl[1-2\frac{G_{\varepsilon}\tau^{-\varepsilon/2}%
}{\varepsilon}\bigl(3g_{1}g_{2}+g_{2}g_{2}^{\prime}\bigr)\bigr]g_{2}^{\prime
}\,,\\
&  \Gamma_{3,0}=g_{0}-(3d)\nonumber\\
&  \qquad=g_{0}-2\frac{G_{\varepsilon}\tau^{-\varepsilon/2}}{\varepsilon
}\bigl[3\bigl(2g_{1}g_{2}+g_{2}g_{2}^{\prime}\bigr)g_{0}\nonumber\\
&  \qquad\qquad\qquad\qquad-\bigl(7g_{1}^{2}+9g_{1}g_{2}^{\prime}%
+3g_{2}^{\prime2}\bigr)g_{1}\bigr]\,.
\end{align}
where all quantities, vertex functions, control parameters, and couplings, are
bare quantities. Recall from the main text that we switch notation when we apply our renormalization scheme in that we put an overcirc over bare quantities, e.g., $\Gamma_{1,0}\rightarrow\mathring{\Gamma}_{1,0}$, and we understand quantities without an overcirc as renormalized ones once the renormalization scheme has been applied. Keeping this in mind when we compare the vertex generating function in its bare and renormalized forms,
\begin{equation}
\Gamma=\sum_{\tilde{k},k}\mathring{\Gamma}_{\tilde{k},k}\frac{\mathring
{\tilde{\varphi}}^{\tilde{k}}\mathring{\varphi}^{k}}{\tilde{k}!k!}%
=\sum_{\tilde{k},k}\Gamma_{\tilde{k},k}\frac{\tilde{\varphi}^{\tilde{k}%
}\varphi^{k}}{\tilde{k}!k!} \, ,%
\end{equation}
we obtain the following renormalizations of the vertex functions.%
\begin{align}
\Gamma_{1,0}  &  =Z^{1/2}\mathring{\Gamma}_{1,0}\,,\\
\Gamma_{1,1}  &  =Z\mathring{\Gamma}_{1,1}\,,\quad\Gamma_{2,0}=Z\Big(\mathring
{\Gamma}_{2,0}+2K\mathring{\Gamma}_{1,1}\Big)\,,\\
\Gamma_{1,2}  &  =Z^{3/2}\mathring{\Gamma}_{1,2}\,,\quad\Gamma_{2,1}%
=Z^{3/2}\Big(\mathring{\Gamma}_{2,1}+2K\mathring{\Gamma}_{1,2}\Big)\,,\\
\Gamma_{3,0}  &  =Z^{3/2}\Big(\mathring{\Gamma}_{3,0}+3K\mathring{\Gamma
}_{2,1}+3K^{2}\mathring{\Gamma}_{1,2}\Big)\,.
\end{align}
Further exploiting the renormalization scheme (\ref{Reno}) and using the scaling-invariant
coupling constants $u=u_{2}u_{2}^{\prime}$, $v=u_{1}u_{2}$, $w=u_{0}u_{2}^{3}$,
it is simple algebra to find
\begin{equation}
Z=1+\frac{u+4v}{6\varepsilon}\,,\quad K=\frac{w-2uv-3v^{2}}{6u_{2}^{2}}\,,
\end{equation}%
\begin{equation}
\underline{\underline{Z}}=\underline{\underline{1}}+\frac{1}{\varepsilon}%
\begin{pmatrix}
2u+4v\,, & 5(w-2uv-3v^{2})/3u_{2}^{2}\\
-u_{2}^{2}\,, & u+4v
\end{pmatrix}
\,,
\end{equation}%
\begin{equation}
\underline{\underline{A}}=\frac{1}{\varepsilon u_{2}}%
\begin{pmatrix}
0\,, & u_{2}^{2}\\
u_{2}^{2}\,, & -2v
\end{pmatrix}
\,,
\end{equation}%
\begin{align}
B_{0}  &  =\frac{11uw+22vw-10u^{2}v-29uv^{2}-22v^{3}}{2\varepsilon}u_{2}%
^{-3}\,,\\
B_{1}  &  =\frac{16uv+39v^{2}-5w}{6\varepsilon}u_{2}^{-1}\,,\\
B_{2}  &  =\frac{2u+6v}{\varepsilon}u_{2}=\bigl(Z_{2}-1\bigr)u_{2}\,,
\end{align}
for the $1$-loop renormalizations.

\section{$2$-loop results of the RG functions}
\label{app:2-loopResults}

Here, we list  our 2-loop results for the RG functions that went into the calculation of the critical exponents for the $\theta$-transition. Details of the calculation leading to these results will be presented elsewhere \cite{JaSt-prep}.

 The $2$-loop parts of the $\alpha$-matrix are given by
\begin{align}
\alpha_{1,1}^{(2)}  &  =1\,,\\
\alpha_{1,2}^{(2)}  &  =-\Big(\frac{47}{24}u+\frac{35}{6}v\Big)\,,\\
\alpha_{2,2}^{(2)}  &  =\Big(\frac{23}{4}u+\frac{161}{12}v\Big)v-\frac{23}%
{12}w\,.
\end{align}
The $2$-loop parts of the $\gamma$- and $\gamma^{\prime}$-function read
\begin{align}
\gamma^{(2)}  &  =\Big(\frac{37}{216}u^{2}+\frac{7}{6}uv+\frac{191}{108}%
v^{2}\Big)-\frac{13}{108}w\,,\\
\gamma^{\prime(2)}  &  =\Big(\frac{29}{72}u+\frac{25}{27}v\Big)w\nonumber\\
&  \quad-\Big(\frac{7}{12}u^{2}+\frac{469}{216}uv+\frac{17}{9}v^{2}\Big)v\,.
\end{align}
The parts of the $\kappa$-matrix that are not given by the shift-invariance:%
\begin{align}
\kappa_{1,1}^{(2)}  &  =\frac{611}{108}w-\Big(\frac{1519}{108}u^{2}%
+\frac{1403}{18}uv+\frac{10873}{108}v^{2}\Big)\,,\\
\kappa_{2,1}^{(2)}  &  =\Big(\frac{43}{3}u^{2}+\frac{3001}{54}uv+\frac{452}%
{9}v^{2}\Big)v\nonumber\\
&  \quad-\Big(\frac{161}{18}u+\frac{580}{27}v\Big)w\,.
\end{align}
The $\beta$-function that is not given by shift-invariance reads
\begin{align}
\beta_{w}^{(2)}  &  =\Big(\frac{55}{2}u^{3}+\frac{10727}{72}u^{2}v+\frac
{4657}{18}uv^{2}+\frac{887}{6}v^{3}\Big)v\nonumber\\
&  \quad-\Big(\frac{2809}{72}u^{2}+\frac{1754}{9}uv+\frac{1391}{6}v^{2}%
-\frac{85}{9}w\Big)w\,.
\end{align}%

\end{document}